\PassOptionsToPackage{table}{xcolor}
\documentclass[manuscript,screen,nonacm]{acmart}
\usepackage{datetime2}
\usepackage{listings}
\usepackage{multirow}
\usepackage{multicol}
\usepackage{makecell}
\usepackage{bigstrut}
\usepackage{siunitx}
\usepackage{tabularx}
\usepackage{xspace}
\usepackage{subcaption}
\usepackage{cleveref}
\begin{document}
\newcommand{\literal}[1]{\texttt{\detokenize{#1}}}
\newcommand{\nvqlink}{\textsc{NVQLink}\xspace}
\newcommand{\defn}[2]{%
  \begin{list}{$\triangleright$}{\leftmargin=3em\itemindent=0em\labelsep=0.5em}%
    \item \textbf{#1} #2%
  \end{list}%
}

% A tame, readable C++ style
\lstdefinestyle{acmcpp}{
  language=C++,
  basicstyle=\ttfamily\footnotesize,
  keywordstyle=\bfseries\color{blue!60!black},
  commentstyle=\itshape\color{green!40!black},
  stringstyle=\color{orange!60!black},
  numbers=left,
  numberstyle=\scriptsize\color{gray},
  numbersep=8pt,
  stepnumber=1,
  breaklines=true,
  columns=fullflexible,
  keepspaces=true,
  showstringspaces=false,
  frame=single,
  tabsize=2,
}
\lstset{style=acmcpp}
% \newcommand{\codeblock}[4]{
% \begin{lstlisting}[
%     language=C++,
%     label={#1},
%     caption={#2},
%     float={#3}
% ]
% #4
% \end{lstlisting}
% }

% Alternate white-gray row highlighting
\rowcolors{2}{}{lightgray}

\lstdefinestyle{mlir}{
    basicstyle=\ttfamily\small,
    keywordstyle=\color{purple}\bfseries,
    commentstyle=\color{green!60!black},
    numbers=left,
    numberstyle=\tiny\color{gray},
    frame=single,
    breaklines=true,
    tabsize=2,
    showstringspaces=false,
    captionpos=b
}
\lstdefinestyle{cudaq}{
    language=C++,
    basicstyle=\ttfamily\small,
    keywordstyle=\color{blue}\bfseries,
    commentstyle=\color{green!60!black},
    stringstyle=\color{red},
    numbers=left,
    numberstyle=\tiny\color{gray},
    frame=single,
    breaklines=true,
    tabsize=2,
    showstringspaces=false,
    captionpos=b
}

\title[Tight Coupling HPC with QPUs]%
  {Platform Architecture for Tight Coupling of High-Performance Computing with Quantum Processors}

\author{Shane A. Caldwell}
% \authornote{These authors contributed equally to this research.}
\email{scaldwell@nvidia.com}
\orcid{0000-0003-3209-8295}
\author{Moein Khazraee}
\orcid{0000-0002-2754-1080}
\author{Elena Agostini}
\orcid{0000-0003-2121-1176}
\author{Tom Lassiter}
\orcid{0009-0006-1019-1228}
\author{Corey Simpson}
\orcid{0009-0005-2063-5825}
\author{Omri Kahalon}
\orcid{0009-0002-4243-7553}
\author{Mrudula Kanuri}
\orcid{0009-0008-3273-4486}
\author{Jin-Sung Kim}
\orcid{0000-0003-2892-8022}
\author{Sam Stanwyck}
\orcid{0009-0004-7391-0088}
\author{Muyuan Li}
\orcid{0000-0003-2035-5135}
\author{Jan Olle}
\orcid{0000-0003-3338-5130}
\author{Christopher Chamberland}
\orcid{0000-0003-3239-5783}
\author{Ben Howe}
\orcid{0009-0002-5372-8890}
\author{Bruno Schmitt}
\orcid{0000-0003-4376-3422}
\author{Justin G. Lietz}
\orcid{0000-0002-8398-5524}
\author{Alex McCaskey}
\orcid{0000-0002-0745-3294}
\affiliation{%
  \institution{NVIDIA Corporation}
  \city{Santa Clara}
  \state{California}
  \country{USA}
}

% A*STAR
\author{Jun Ye}
\orcid{0000-0003-1963-0865}
\affiliation{
    \institution{Institute of High Performance Computing (IHPC), Agency for Science, Technology and Research (A*STAR)}
    \city{Singapore}
    \state{Singapore}
    \country{Singapore}
}

% PNNL
\author{Ang Li}
\orcid{0000-0003-3734-9137}
\affiliation{%
  \institution{Pacific Northwest National Laboratory}
  \city{Richland}
  \state{Washington}
  \country{USA}
}
\affiliation{%
  \institution{University of Washington}
  \city{Seattle}
  \state{Washington}
  \country{USA}
}

% Sandia QPL
\author{Alicia B. Magann}
\orcid{0000-0002-1402-3487}
\author{Corey I. Ostrove}
\orcid{0000-0001-8274-3331}
\author{Kenneth Rudinger}
\orcid{0000-0002-3038-4389}
\author{Robin Blume-Kohout}
\orcid{0000-0001-8134-948X}
\author{Kevin Young}
\orcid{0000-0002-4679-4542}
\affiliation{%
  \institution{Quantum Performance Laboratory, Sandia National Laboratories}
  \city{Albuquerque}
  \state{New Mexico}
  \country{USA}
}

% MITLL
\author{Nathan E. Miller}
\orcid{0000-0003-1523-6553}
\affiliation{%
 	\institution{Lincoln Laboratory, Massachusetts Institute of Technology}
 	\city{Lexington}
  	\state{Massachusetts}
 	\country{USA}
}

%LBNL
\author{Yilun Xu}
\orcid{0000-0001-9284-809X}
\author{Gang Huang}
\orcid{0000-0002-3249-9315}
\affiliation{%
  \institution{Lawrence Berkeley National Laboratory}
  \city{Berkeley}
  \state{California}
  \country{USA}
}
% LBNL + Cal
\author{Irfan Siddiqi}
\orcid{0000-0003-2200-1090}
\affiliation{%
  \institution{University of California, Berkeley}
  \city{Berkeley}
  \state{California}
  \country{USA}
}
\affiliation{%
  \institution{Lawrence Berkeley National Laboratory}
  \city{Berkeley}
  \state{California}
  \country{USA}
}

% ORNL
\author{John Lange}
\orcid{0000-0003-0616-7437}
\affiliation{
  \institution{Oak Ridge National Laboratory}
  \city{Oak Ridge}
  \state{Tennessee}
  \country{USA}
}
\affiliation{
  \institution{University of Pittsburgh}
  \city{Pittsburgh}
  \state{Pennsylvania}
  \country{USA}
}
\author{Christopher Zimmer}
\orcid{0000-0001-5054-4354}
\author{Travis Humble}
\orcid{0000-0002-9449-0498}
\affiliation{
  \institution{Oak Ridge National Laboratory}
  \city{Oak Ridge}
  \state{Tennessee}
  \country{USA}
}

\renewcommand{\shortauthors}{Caldwell, McCaskey, et al.}

\begin{abstract}
    We propose an architecture, called \nvqlink, for connecting high-performance computing (HPC) resources to the control system of a quantum processing unit (QPU) to accelerate workloads necessary to the operation of the QPU.
    We aim to support every physical modality of QPU and every type of QPU system controller (QSC).
    The HPC resource is optimized for real-time (latency-bounded) processing on tasks with latency tolerances of tens of microseconds.
    The network connecting the HPC and QSC is implemented on commercially available Ethernet and can be adopted relatively easily by QPU and QSC builders, and we report a round-trip latency measurement of \qty{3.96}{\micro\second} (max) with prospects of further optimization.
    We describe an extension to the CUDA-Q programming model and runtime architecture to support real-time callbacks and data marshaling between the HPC and QSC.
    By doing so, \nvqlink extends heterogeneous, kernel-based programming to the QSC, allowing the programmer to address CPU, GPU, and FPGA subsystems in the QSC, all in the same \literal{C++} program, avoiding the use of a performance-limiting HTTP interface.
    We provide a pattern for QSC builders to integrate with this architecture by making use of multi-level intermediate representation dialects and progressive lowering to encapsulate QSC code.
\end{abstract}


\maketitle

\section{Introduction}\label{scn:intro}

Quantum computing is one of the most promising technologies of our era, and its invention is a great challenge in science and engineering.
There is a worldwide, multi-decade research effort going on to meet that challenge.
While there is much exciting progress in that effort, today's quantum processing units (QPUs) are somewhat wild objects belonging to an industry engaged in deep and rapidly evolving research.
Many uncertainties must be resolved on our way to a robust and compelling example of useful quantum computation.

\subsection{Two regimes of HPC integration}\label{scn:hpc-regimes}
It is widely anticipated, though not a position of this paper, that unlocking the potential of quantum computing will require integrating the QPU within the traditional high-performance computing (HPC) environment of a supercomputer~\cite{Mohseni-2024-HowBuildQuantum-Preprint, Burgholzer-2025-MunichQuantumSoftware-Preprint, Mansfield-2025-FirstPracticalExperiences-Preprint,buchs2025role}.
In this model, the QPU acts as a specialized node within the supercomputer.
The QPU augments the supercomputer, and we may think of the new machine as a quantum-accelerated supercomputer.
There may be many quantum nodes in such a machine, so as to implement circuit knitting or similar protocols to divide work among the nodes.
The QPUs may also be linked to each other in a quantum network to distribute entanglement, in which case their behavior may be like a larger quantum computer with its own heterogeneous quantum resources~\cite{Cisco-2025-QuantumDataCenter-Preprint}.
The supercomputer is expected to be running its own compute-intensive part of the application, and it may be free to treat the QPU as a black box.
If the QPU is operating with some combination of quantum error correction (QEC), quantum error mitigation (QEM), and quantum ensemble sampling, the data throughput and latency requirements between the supercomputer and the QPU may be very modest relative to typical node-to-node interconnects in the supercomputer.

We expect a second type of HPC integration will be required, where the HPC resource must be \emph{tightly coupled} to the control system of the QPU to perform online workloads of the QPU itself.
In this scenario, which is the subject of this paper, the HPC resource is probably of a smaller scale than a supercomputer and is optimized for real-time computation in the critical path of QPU operation, serving functionally as a part of the QPU.
The best known of these workloads is QEC decoding, which in several QPU modalities requires the lowest possible latency while sustaining network throughput of $\sim\qty{1}{Tb/s}$ and compute of up to $\sim\qty{1}{PFLOP/s}$~\cite{NERSC-2024-EvaluationClassicalHardware}.
We expect QPUs will also require continual recalibration to keep quantum operational fidelities within tolerances, and the computational demands of calibration procedures may rise as tolerances shrink, as longer-range errors have to be suppressed~\cite{Klimov-2024-OptimizingQuantumGates}, and as machine learning is increasingly used to improve automation and accuracy.
These tasks require specialized hardware and programming that are not present in the supercomputing environment, and they rightly belong to the QPU as conditions of its operation.

\subsection{Requirements from QEC}\label{scn:qec-requirements}
Tight coupling is critical for QEC decoding.
Consider a QPU running a program encoded with a stabilizer code.
The code's stabilizer measurements produce a continuous stream of \emph{syndrome} data emitted by the QSC at the QEC cycle rate, which is up to $\sim \qty{1}{\mega\hertz}$ in systems planned today~\cite{tourdegross, GoogleQuantum-2024-QuantumErrorCorrection, Quantinuum-2021-DemonstrationTrappedionQuantumCCD, Quera-2024-LogicalQuantumProcessor, Infleqtion-2025-UniversalNeutralAtomQuantum, Diraq-2017-SiliconCMOSArchitecture, Intel-2022-SpiderwebArraySparse, PsiQuantum-2023-FusionbasedQuantumComputation}.
Each syndrome measurement is 1 bit per ancilla qubit per QEC cycle, though this would increase to perhaps 16 bits per ancilla in the case that continuously varying (``soft") readout data is used as input~\cite{Pattison-2021-ImprovedQuantumError-Preprint, IBMQuantum-2024-SoftInformationDecoding, AlphaQubitDecoder, Akahoshi-2025-RuntimeReductionLattice-Preprint}.
The QEC decoder must process all of the syndrome data to maintain a record of the current state of the QPU, and information from this record is accessed for feedforward operations at specific points in the program.
A useful and common paradigm for determining the compiled structure of such a program on a topological code is found in \emph{Pauli-based computation}~\cite{IBMQuantum-2016-TradingClassicalQuantum} with \emph{lattice surgery}~\cite{Horsman-2012-SurfaceCodeQuantum} and compilation into a sequence of concurrent Pauli measurements with non-Clifford gates at each layer~\cite{Litinski2019gameofsurfacecodes}.

There are two timing requirements on the decoder performance:
\begin{enumerate}
\item Decoder throughput, measured as syndrome cycles decoded per unit of time, must keep up with the continuous stream of syndrome data.
At each feedforward event, the decoder's record will be delayed from the stabilizer rounds by some amount of real time and some number of stabilizer cycles which constitute a backlog for the decoder.
If syndrome processing falls short of the streaming rate over one feedforward period, the backlog grows proportionally to that shortfall which leads to exponential growth of the shortfall and eventual stalling of the QPU~\cite{terhal2015}.
\item Reaction time, measured as time elapsed between the last syndrome measurement acquisition on the QSC to the application of a feedforward action derived from syndrome data up to that point, directly impacts both QPU clock speed and fidelity.
\end{enumerate}
We emphasize that additional latency in the reaction time impacts the correctness of a quantum program's execution, even in a fault-tolerant quantum processor.
Consider a fault-tolerant QPU that makes effective use of QEC to reduce its per-instruction error probability $\varepsilon$ by many orders of magnitude below the error rates of its physical operations.
A program on this QPU will be free of any logical error with probability $P = (1-\varepsilon)^N$, implying an exponential cutoff in success probability at a program of size $N_{\text{limit}} \sim 1/\varepsilon$.
The same argument applies if we count only instructions which perform non-Clifford gates and require feed-forward events in the QEC protocol.
While waiting for the feed-forward operation, the target logical qubit can run syndrome extraction cycles to remain coherent, and a key metric of interest in QEC is the logical error rate per cycle.
This means that the logical error incurred by the reaction time can be estimated with the logical error rate per cycle along with the number of cycles spent idle.
At these events, which are also used to estimate program runtimes (for example in~\cite{Gidney-2025-HowFactor2048-Preprint}), the instruction error probability increases linearly with respect to the feed-forward reaction time of the QEC decoding system~\cite{Chamberland-2022-UniversalQuantumComputing, Chamberland-2023-TechniquesCombiningFast-Preprint, Skoric-2023-ParallelWindowDecoding}.
So we see that system latencies affect not only the rate of processing speed but the tradeoff between a program's execution time and the correctness of its execution.

Decoder throughput and reaction time are similar to what is seen in superscalar processors that contain multiple execution units in a single core~\cite{Smith-1995-MicroarchitectureSuperscalarProcessors}.
Due to parallel resources (execution units), the total number of instructions executed per second is not simply the inverse of the end-to-end execution time of a single instruction.
In this analogy, the decoder throughput (syndrome extraction cycles decoded per second) is analogous to the number of instructions retired per second, and the reaction time is analogous to the end-to-end instruction execution time.

There is currently no experimentally demonstrated solution for QEC decoding at scale, ie. in systems that effectively operate $\sim 100$ logical qubits or more in a fault tolerant quantum program containing $\sim 1$ million logical instructions.
There is very active ongoing research in QEC codes as well as algorithmic~\cite{panteleev2021degenerate, wu2023fusion, higgott2025sparse, Chan-2024-SnowflakeDistributedStreaming-Preprint, Hillmann-2024-LocalizedStatisticsDecoding-Preprint, Ziad-2024-LocalClusteringDecoder-Preprint, Gong-2024-LowlatencyIterativeDecoding-Preprint, beni2025tesseractsearchbaseddecoderquantum, muller2025relayBP, Wang-2025-FullyParallelizedBP-Preprint, Wolanski-2025-AmbiguityClusteringAccurate-Preprint, Zhang-2025-LATTEDecodingArchitecture-Preprint, Turner-2025-ScalableDecodingProtocols-Preprint, Riverlane-2025-RealtimeScalableFast-Preprint}
and AI-based~\cite{AlphaQubitDecoder, Blue-2025-MachineLearningDecoding-Preprint, Lange-2025-DatadrivenDecodingQuantum, Mi-2025-UncertaintyAwareGeneralizableNeural-Preprint, Hu-2025-EfficientUniversalNeuralNetwork-Preprint, Cao-2023-QecGPTDecodingQuantum-Preprint, Liu-2025-DecodingQuantumLow-Preprint, Varbanov-2025-NeuralNetworkDecoder, Australia3DConv} decoders.
Today's state of the art in applied QEC comprises early demonstrations of partial functionality:
single-qubit state preservation including below-threshold~\cite{GoogleQuantum-2024-QuantumErrorCorrection} and break-even~\cite{AWSQuantum-2025-HardwareefficientQuantumError, Sivak-2023-RealtimeQuantumError, Brock-2025-QuantumErrorCorrection} demonstrations;
specialized demonstrations of quantum logic on a few logical qubits~\cite{Quantinuum-2021-RealizationRealTimeFaultTolerant, Quantinuum-2024-DemonstrationLogicalQubits-Preprint, Quantinuum-2025-BreakingEvenMagic, Quantinuum-2025-QuantumErrorCorrectedComputation-Preprint, Quera-2024-LogicalQuantumProcessor, Infleqtion-2024-FaultTolerantOperationMaterials-Preprint, Infleqtion-2025-DemonstrationLogicalArchitecture-Preprint, AtomComputing-2025-FaulttolerantQuantumComputation-Preprint, AtomComputing-2025-RepeatedAncillaReuse-Preprint};
and demonstrations of decoding on FPGA and CPU with sufficient throughput on small codes~\cite{Caune-2024-DemonstratingRealtimeLowlatency-Preprint, GoogleQuantum-2024-QuantumErrorCorrection,ibmFPGAdecoder}.
A number of analyses of requirements for the quantum system controller are available~\cite{Battistel-2023-RealTimeDecodingFaultTolerant, QuantumMachines-2024-ControllerdecoderSystemRequirements-Preprint, QuantumMachines-2024-BenchmarkingAbilityController-Preprint}, and a viewpoint on the tradespace of specialized QEC decoding hardware (GPU, FPGA, ASIC) is available in Ref.~\cite{riverlane2024qechardware}.
Architectures based entirely in firmware (FPGA) or hardware (ASIC), where decoding graphs are physically instantiated in hardware, contemplate purpose-built networks for distributed computation at ultralow latency~\cite{tourdegross, liyanage2024fpga}.

To achieve fault tolerant quantum computing at scale, the throughput requirement on QEC decoding will necessitate heavy use of parallel processing.
For example, a fault-tolerant algorithm using lattice surgery with topological codes requires operations with very large effective code distances. In addition to processing multiple syndrome measurement rounds for all logical qubits, large code patches need to be processed per round in parallel over both space and time~\cite{Skoric-2023-ParallelWindowDecoding,AlibabaParallelWindow}. By partitioning the lattice into blocks of spatial dimensions $\mathcal{O}(d)$ (for a distance $d$ code) and temporal dimensions $\mathcal{O}(d_m)$ (where $d_m$ is the number of syndrome measurement rounds), all such blocks can be decoded in parallel. In~\autoref{scn:scalable-ftq-programs} we discuss such requirements in more detail to achieve scalable real-time decoding algorithms. We also discuss the important role GPUs can play for block-wise decoding, especially when using AI-based pre-decoders~\cite{Chamberland-2023-TechniquesCombiningFast-Preprint,Australia3DConv}.

In contrast to topological codes, quantum low-density parity check (qLDPC) codes such as bivariate bicycle codes~\cite{bravyi2024high} offer lower qubit overhead.
The modular decoding structure and parallelization strategies for topological codes cannot directly be ported to qLDPC decoding, but novel decoding strategies are actively being investigated~\cite{tourdegross, muller2025relayBP, ibmFPGAdecoder}.

There are a number of practical considerations that favor a heterogeneous architecture that admits accelerated computing hardware and a programming model that support rapid iteration.
\begin{enumerate}
    \item Within the execution of one fault-tolerant quantum program, it is necessary to dynamically refer to the relevant detector error matrix (DEM)~\cite{Derks-2024-DesigningFaulttolerantCircuits-Preprint} for the logical operation at hand.
    Logical gates require dynamic reconfigurations of their Tanner graph, and each \emph{applied} 
    logical gate (ie. logical gate as applied to particular logical qubits in the QPU) requires its own DEM values for decoding.
    Thus, strategies for handling dynamical changes to DEM structure and values, or DEM modularization, are likely necessary.
    \item QPUs are subject to physical drift, such that the noise model informing the DEM changes over time.
    The decoder configuration must be maintained continuously to preserve high logical fidelity.
    \item In a high distance code, it will generally be impossible to align the readout hardware with every DEM in the needed set of DEMs.
    Therefore a network and syndrome aggregation mechanism are required.
    \item The rate of research in the QEC space is such that many researchers and builders are likely to innovate on QEC codes and decoders and want to share solutions.
    An open platform supporting such sharing and shortened time to solution for new implementations will be valuable.
\end{enumerate}

The total compute required for decoding at scale can be estimated from Figure 10 of Ref.~\cite{NERSC-2024-EvaluationClassicalHardware}, which finds that a 100-qubit QPU can run a depth-$10^6$ program encoded by a surface code if it has Fusion Blossom~\cite{Wu-2023-FusionBlossomFast, Wu_fusion-blossom_2023} running at a compute intensity of \qty{200}{\tera FLOP/s}.
Fusion Blossom at \qty{1}{\peta FLOP/s} supports a depth-$10^9$ program on a 1000-qubit QPU.
If we instead assume the use of an AI decoder, we would estimate needing \qty{2}{FLOP} per logical qubit, per model parameter, per inference.
Assuming we use a 25-million parameter model ($5\times$ the size of the AlphaQubit model~\cite{AlphaQubitDecoder}) and decode at a \qty{1}{\mega\hertz} QEC cycle, we would need \qty{50}{\tera FLOP/s} per logical qubit to run the decoder.
We may add a factor of ten to this to ensure we have headroom for dynamic compilation and realtime updating of QPU control parameters and the decoder noise model and conclude that \qty{50}{\peta FLOP/s} provides a conservative estimate for the necessary compute intensity at the scale of 100 logical qubits.

\subsection{Challenges}\label{scn:tight-coupling-challenges}

Introducing tightly coupled HPC to the QPU environment presents a number of challenges that have emerged as priorities within the quantum-HPC community, which has started seeking solutions that scale up with increasing capabilities of quantum computing systems~\cite{openqse2025}.
Most QPU system controllers (QSCs) are implemented using field programmable gate array (FPGA) or radio-frequency system-on-chip (RFSoC) devices, which run a firmware-defined pulse processor unit (PPU)~\cite{xu2021qubic,xu2023qubic}.
(Please look ahead to~\autoref{fig:qpu} for a system diagram identifying components discussed here.)
There are dozens of PPU implementations in use in the quantum computing industry, most of which are closed-source and proprietary.
Each PPU has its own features, instruction set, and compiler toolchain that lowers from programs expressed in gate-level or pulse-level quantum instruction languages such as CUDA-Q, QASM, Cirq, Quil, and others, as well as lower-level languages that introduce control over pulse scheduling, waveform definitions, and other digital signal processing features common in modern software-defined radio technology.
The whole QSC system integrates an array of PPUs with state-of-the-art synchronization of all PPU clocks (\qty{250}{\mega\hertz} typ.), microwave phase coherence, and often an ultra-low latency $\sim\qty{300}{\nano\second}$ network among the PPUs.

Many features and requirements of the QSC differ among the QPU modalities as well as the various implementations available from QSC vendors.
To give several examples:
\begin{itemize}
    \item Some modalities achieve all-to-all qubit connectivity within the QPU by dynamically shuttling trapped ions or atoms in space, either to interact with neighbors or to enter zones of the trap where specific control and readout functions are performed.
    Other modalities have fixed qubit topologies, and no dynamic routing constraints impact the PPU schedule.
    \item QPU modalities differ in the presence or absence of optical carrier waves for control and readout.
    \item QSCs have variable level of dynamism and programmability.
    Some QSCs support only static sequences or nonparameterized arbitrary waveforms, while others support fully dynamic pulse pipelines and real-time-patchable parameterized waveforms.
    \item QSCs have variable level of sophistication in compiler toolchain.
    The ability to mix gate-level and pulse-level programming paradigms, and to mingle runtime concerns such as QEC encoding, is still in a nascent stage and not standardized.
    \item QSCs have variable network topologies and sophistication of pulse scheduling functionality within the QSC.
    Some QSCs may support a global pulse scheduler that maintains resolution of every clock tick of every PPU, while others do not.
    \item Physical qubit state readout mechanisms are very different from one modality to another, which entails very different hardware scaling properties with respect to physical qubit counts and very different data reduction procedures that are typically encoded in PPU firmware.
    \item Some QSCs will place control and readout electronics into the physical QPU environment, including deep cryogenic environments with extreme power limitations~\cite{Liu-2023-SingleFluxQuantumBased, IBMQuantum-2023-UsingCryogenicCMOS-Preprint, Bartee-2025-SpinqubitControlMillikelvin}.
\end{itemize}

\subsection{Goals and outline}\label{scn:goals}
To summarize, our view is that a program defined on a QPU is part of a program defined on a supercomputer.
The quantum portion of the program must be compiled down through quantum instructions and eventually to pulses scheduled across every PPU in the QSC, taking on sensitivity to the PPU architecture, the physical QPU modality and instance, and the current calibrated state of the physical QPU.
Yet from within the PPU program running in the we must send data to a tightly coupled resource for QEC decoding.
We want to be able to write everything from the parent application and the body of the decoding function in common programming languages such as Python, \literal{C++}, and CUDA.
We want the resulting program to be orchestrated on a distributed real-time system that does not rely on the use of web-oriented networking technology such as HTTP transport.
And we want to do this while allowing every QSC builder to achieve this integration with minimal effort and changes to their toolchain.

By defining the \nvqlink architecture, we aim to advance the integration of quantum processors into the supercomputing environment by the following means.
\begin{enumerate}
    \item Introducing an industrial grade, real-time (deterministic latency) interface between a QSC and HPC resources for online quantum error correction and all other compute-intensive online workloads of the QPU.
    \item Providing programming mechanisms and efficient data marshaling among many heterogeneous devices within a Logical QPU.
    \item Defining a platform that scales from current-generation QPU devices to future fault-tolerant systems.
    \item Enabling offline program development and validation through substitutable PPU emulation (VPPU) to reduce dependencies on physical hardware.
    \item Offering steps toward standardization in parts of the system architecture where innovation is not needed or desired.
    \item Doing the above in a way that enables all QPU and QSC builders to take advantage of the platform to advance their own successes.
\end{enumerate}

In \autoref{scn:system} we describe the hardware architecture of a Logical QPU.
In \autoref{scn:programming-model} we introduce proposed CUDA-Q extensions for real-time device callbacks and heterogeneous memory.
In \autoref{scn:runtime} we propose a new trait-based runtime, compiled-kernel format, executors, and a Logical QPU Driver API.
In \autoref{scn:spec} we suggest preliminary requirements in an effort to work in the open and gather feedback from interested readers.
In \autoref{scn:qpu-workloads} we describe in greater detail the QPU-level workloads that we expect will benefit from tight coupling.
In \autoref{scn:development} we discuss the use of development and simulation tool (PPU emulation, Physical QPU simulation, and QEC simulation).

\section{System Architecture}\label{scn:system}

\begin{figure}[ht]
    \centering
    \includegraphics[width=0.9\linewidth]{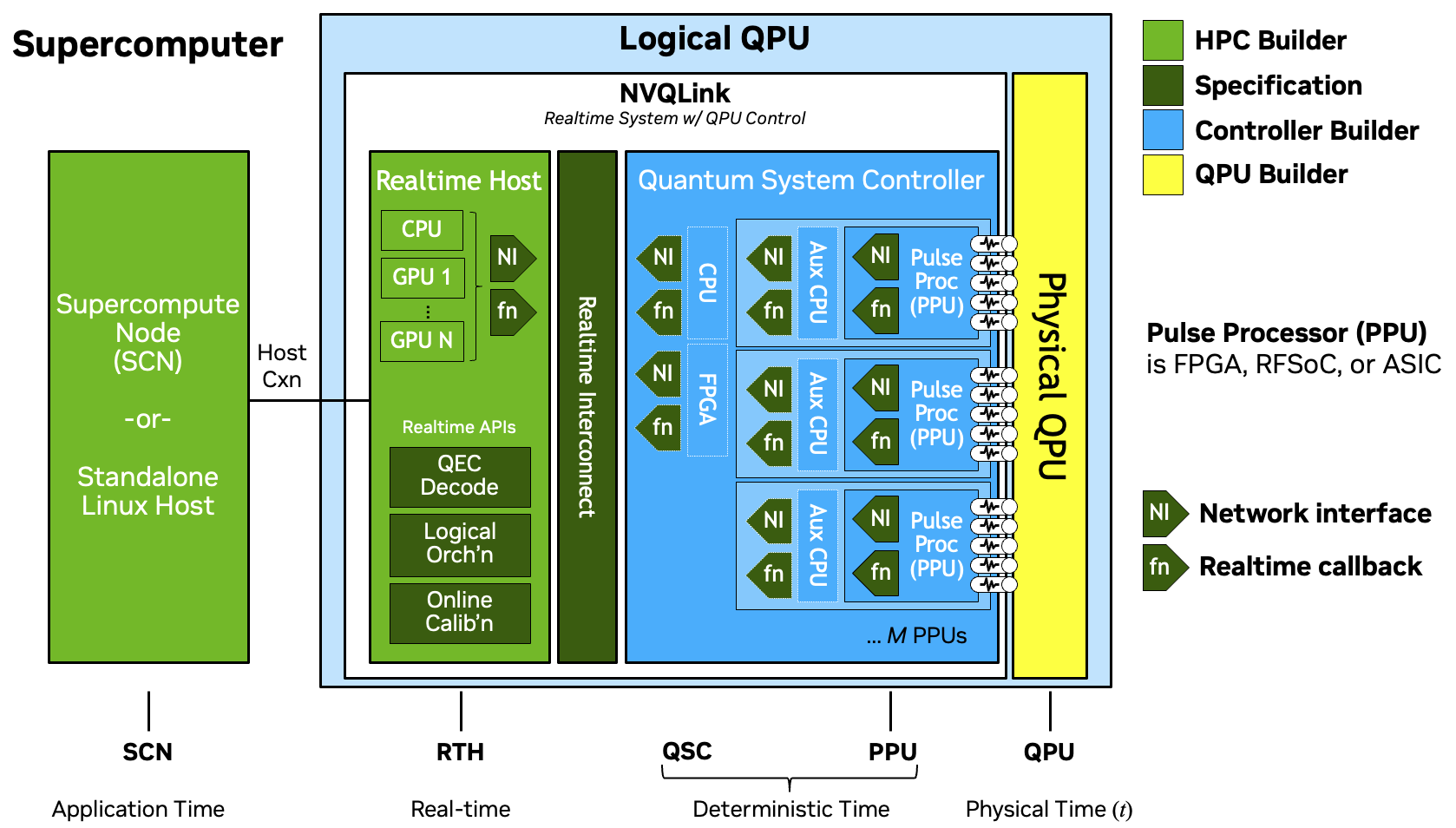}
    \caption{
        Machine model of the Logical QPU.
        The \nvqlink architecture comprises the Real-time Host (RTH) and QPU Control System (QSC) connected by a low-latency, scalable Real-time Interconnect joining them into a network capable of handling the runtime workloads of a Fault Tolerant Quantum Computer.
        The RTH contains traditional HPC compute resources (CPUs and GPUs and perhaps other specialized hardware), while the QSC contains the Pulse Processing Units controlling the QPU.
        These compute resources also comprise the key memory and storage resources for the application to consider and orchestrate including GPU memory, RTH system main memory, etc.
        The programming model for this system is built to recognize all CPUs, GPUs, and resources within the QSC including CPUs, PPUs, and other specialized FPGA resources, as targetable 
        Devices and enable Real-time Callback functions (\literal{fn}) among them to support distributed processing and data marshaling.
        To support this, a small and optional Network Interface (NI) is provided that enables unilateral and private adoption by the QSC builder.
        This construction affords flexibility in the value chain of Physical QPU, QSC, and runtime protocols such as QEC Decoding and Online Calibration: each of these components is provided by a third party who may build or integrate some or all of the system and who may share their implementation of each component or keep it proprietary at their discretion.
        The purpose of this architecture is support such flexibility while enabling every implementation to achieve state-of-the-art HPC performance at minimal cost and time to solution.
    }
    \label{fig:qpu}
\end{figure}

The \nvqlink system comprises the real-time execution environment in which an HPC system is tightly coupled to a QPU control system.
The system is represented in~\autoref{fig:qpu}.

\subsection{System components}\label{scn:system-components}

The \nvqlink architecture defines set of hardware components and their relationships that constitute a \emph{machine model} for tightly coupled HPC and quantum programming.
This machine model corresponds to a subset of the systems that can be defined in the CUDA-Q programming model, which is described in more detail in~\autoref{scn:programming-model}.

\defn{Physical QPU}{
    (PQPU) Quantum system whose physical state is described by a vector in a computational Hilbert space.
    This is the quantum object in the system and the key resource that enables quantum computation.
    We refer to it independently of its control and readout electronics.
    Examples include superconducting electrical circuits, spin qubits, trapped atoms and ions, flying photon arrays, etc.
}
\defn{Logical QPU}{
    (LQPU) System capable of quantum computation, necessarily including the control and readout electronics.
}
\defn{Quantum System Controller}{
    (QSC) System that implements quantum coherent control and readout on the Physical QPU.
    This typically contains an array of PPUs comprising all the analog I/O channels to and from the Physical QPU.
    We are abstracting over the choice of whether to group PPUs into smaller chassis or assemble them into a single-layer array.
}
\defn{Pulse Processing Unit}{
    (PPU) Unit of control and readout electronics.
    Analog control unit containing one or more Pulse Processors, typically implemented on one or more FPGAs.
    The PPUs perform control and readout on physical qubits.
    Modern PPUs typically use Software Defined Radio (SDR) techniques relying on Numerically Controlled Oscillators (NCO) for microwave carrier frequencies.
    \autoref{fig:qpu} identifies four components of the PPU
    \begin{enumerate}
    \item Network Interface (NI)
    \item Pulse Processor (typically implement in FPGA or RFSoC firmware)
    \item Analog/Digital Converters (A/D)
    \item Analog Frontend, which comes with electrical isolation effective at microwave frequencies and has analog I/O ports in a format such as SMA.
    \end{enumerate}
    This architecture attempts to be agnostic to all behaviors and implementation details of the PPU except its means of communicating with the Real-time Host.
}
\defn{Virtual Pulse Processing Unit}{
    (VPPU) A PPU emulator that can be substituted for the actual PPU of the
    \nvqlink system for offline development and system simulation.
    VPPU implementations must fully implement the instruction set of the
    target PPU and maintain compiler compatibility.
}
\defn{Real-time Host}{
    (RTH) High-performance computing resource used for workloads in the Real-time Domain (see below).
    This will be a GPU node or multi-node cluster programmable by CUDA-Q, and its CPU a Host in terms of the CUDA Programming Model~\cite{cuda-heterogeneous}.
}
\defn{Real-time Interconnect}{
    Network switch connecting the Real-time Host to the QSC in some way and perhaps to every PPU directly.
    The network architecture of \nvqlink, described further in~\autoref{scn:network}, is optimized for low latency performance on a static network and supporting easy integration into proprietary firmware on the QSC.
}
\defn{Access Node}{
    A traditional (non-real-time) compute node that serves remote clients, communicates with local and online storage, processes requests, etc.
    In a standalone configuration typical of usage by a QPU builder, it is usually a server running a common Linux distribution on a local network with the QSC.
    In a supercomputing integration it may be a node within the supercomputer.
    The only assumption we make on the Access Node is that it should connect to the Real-time Host with performance sufficient to run the application.
}

% Commentary on system components
The primary hypothesis motivating the inclusion of the Real-time Host in the Logical QPU in the \nvqlink model is the expectation that useful quantum computation will require tight coupling.
We envision a future in which the Logical QPU is the \textit{de facto} meaning of the term "QPU," and in CUDA-Q this is the meaning of the kernel attribute \literal{__qpu__}.
The Logical QPU is a heterogeneous device comprising other processors, some of which are also programmable using CUDA.
The Logical QPU itself is programmable using CUDA-Q.

While work is ongoing to realize this vision, we expect the Access Node to continue to connect to the QSC by whatever means exist in prior integrations.
This configuration must be left to the QPU and QSC builder who maintain these prior integrations.
The Real-time Host can be introduced gradually in a time-sharing mode where its functions are brought online with mixed usage.

The QSC is subject to its own set of requirements which vary by Physical QPU modality, architecture, and version.
This architecture is intentionally mute on the QSC's requirements, so that it can vary maximally in response to the needs of the Physical QPU.

\subsection{Time domains}\label{scn:time}

We identify and distinguish four time domains that are required in the system.

\defn{Physical Time Domain}{
    (PTD) The time domain describing the Physical QPU in the lab frame, ie. the continuous variable $t$ in the control Hamiltonian of the PQPU.
}
\defn{Deterministic Time Domain}{
    (DTD) Time domain in which quantum-coherent control and readout operations are scheduled.
    Typically this domain is defined discretized by the synchronized clock of the FPGAs in the QSC.
}
\defn{Real-time Domain}{
    (RTD) Time domain in which tasks are performed that either require data that cannot be localized within the QSC or require more compute intensity than the QSC can support, yet nevertheless require some contract on overall latency.
    The clearest requirements for this domain come from running online quantum error correction, but we anticipate that other valuable uses of the RTD will be found before fault tolerant quantum computing is fully realized.
}
\defn{Application Time Domain}{
    (ATD) Time domain in which the application of the Logical QPU is executed.
    This is in all respects a traditional HPC domain, and the latency of operations affects application performance in ways that are no different from traditional HPC applications.
}

The DTD and RTD operate under real-time contraints due to the latency requirements with which they must produce and return results to the QSC. 
In order for the system to ensure correct operation these constraints can be specified using a set of inputs to the DTD and RTD that define the real-time requirements that the DTD and RTD must meet. 
These constraints will enable the DTD and RTD to determine the satisfiability of the requested operations as well as to ensure that the system schedule will meet the requirements. 
As in other real-time systems we expect these requirments to be expressed either as individual deadlines for each requested operation, or a periodicity requirment that constrains the processing time for a recurring set of operations. 
These inputs to the DTD and RTD will allow the resources to internally manage the workload schedule and provide feedback to indicate a violation of the real-time requirements. 
Additionally, it will enable the DTD and RTD to independently adjust the quality of results to potentially provide a tradeoff between accuracy and timeliness, for instance an operator could return early with 
intermediate/sub-optimal results if its determined that fully computing the operation would result in a missed deadline.

\subsection{Network architecture}\label{scn:network}

Implementations of the Real-time Interconnect in an \nvqlink system are open to third parties with their own requirements.
Therefore we do not stipulate such requirements as a maximum acceptable latency, minimum throughput, or even a topology.

We instead provide open access to a highly performant and scalable reference implementation that aims to satisfy the following criteria.

\begin{enumerate}
\item A standard protocol optimized for a static, point-to-point network.
\item A freely available FPGA IP core implementing the protocol that QSC builders may integrate privately into their proprietary firmwares.
\item Apart from FPGA IP core, reliance only on widely available networking equipment.
\item Latency as low as possible, subject to the constraints above.
\item Throughput and network radix well in excess of current QPU scales.
\end{enumerate}

To achieve this goal, the number of components involved in the communication should be minimized, while scaling the system to a high number of PPUs is desired. Even though putting the QSC system as a Peripheral Component Interconnect Express (PCIe) card can enable direct communication between QSC and GPU, it encounters the scaling challenges of PCIe. Main challenge being limited number of PCIe slots to insert such cards in commodity equipment, as well as challenges of increasing the number of slots. Also from a development point of view, developing such cards for newer versions of PCIe bus, or maintaining the driver and software can be challenging.

The alternative is to use a network interface card (NIC) and connect the QSC to the HPC system through commodity Ethernet or InfiniBand physical links. The increased latency between QSC and NIC from packet generation and going across the transceivers and the physical link is in order of nanoseconds, with minimal jitter. Benefiting from the hardware acceleration within the NIC, the added latency to get to the PCIe bus is also low with low jitter. If desired, Ethernet or InfiniBand switches can be added, which also have accelerated hardware. This alternative enables easier scaling, both in terms of link speed to transfer more data per PPU, as well as use of network switches to scale to more PPUs and aggregating the data. Therefore, we proceeded with this approach, as the added latencies and jitter were acceptable, especially compared to typical higher latency and jitter considerations for the PCIe bus.

Note that the number of components in the processing of the packet is still important and should be minimized. Therefore, we used NIC offload features such as Remote Direct Memory Access (RDMA) to bypass the host processor and kernel, as well as DOCA GPUNetIO library~\cite{gpunetio} to enable direct packet handling from the GPU instead of the host processor. In other words, benefiting from these two technologies, only the NIC and GPU are involved during the processing of packets coming from and going to the QSC, without any host involvement. Also by using an HPC system that has a dedicated PCIe switch between the NIC and GPU slots, the data does not traverse to the host processor socket and stays between the NIC and GPU with a single hop over the switch. This avoids contention from other components on the shared PCIe bus.

Another point to consider is when latency and jitter is the main objective, using a reliable connection becomes less desirable, as that reliability translates into packet retransmission in case of dropped packets due to errors such as checksum error. Such retransmissions usually happen after some timeout, and even if a specific system has features to notify the sender of such a drop, the added latency of this retransmission can throw off the timing of the consecutive packets, with minimal control from the software side. On the other hand, for this specific use case, the size of the data sent per transmission is really small, and using a certified Ethernet connection with low bit error rate (BER), such as NVIDIA's 100G Ethernet cables engineered for a BER of less than \(10^{-15}\), makes the probability of such drops minuscule. 

If detection of such drops are desired, we can include a packet number in each packet and expose it to the software. If a higher reliability is desired, we can replicate the small packets, which achieves a more predictable latency and jitter. For this use case with periodic small packets, typically the bandwidth is not a bottleneck, and if the rate and size of data becomes a bottleneck, a higher speed Ethernet/InfiniBand link can be used. Therefore, we opted for an unreliable connection between the QSC and the NIC, and kept the flexibility of handling such rare occurrences in the hand of the user and software, if necessary.

\subsection{Network proof of concept}\label{scn:network-poc}

\begin{figure}[ht]
    \centering
    \includegraphics[width=0.6\linewidth]{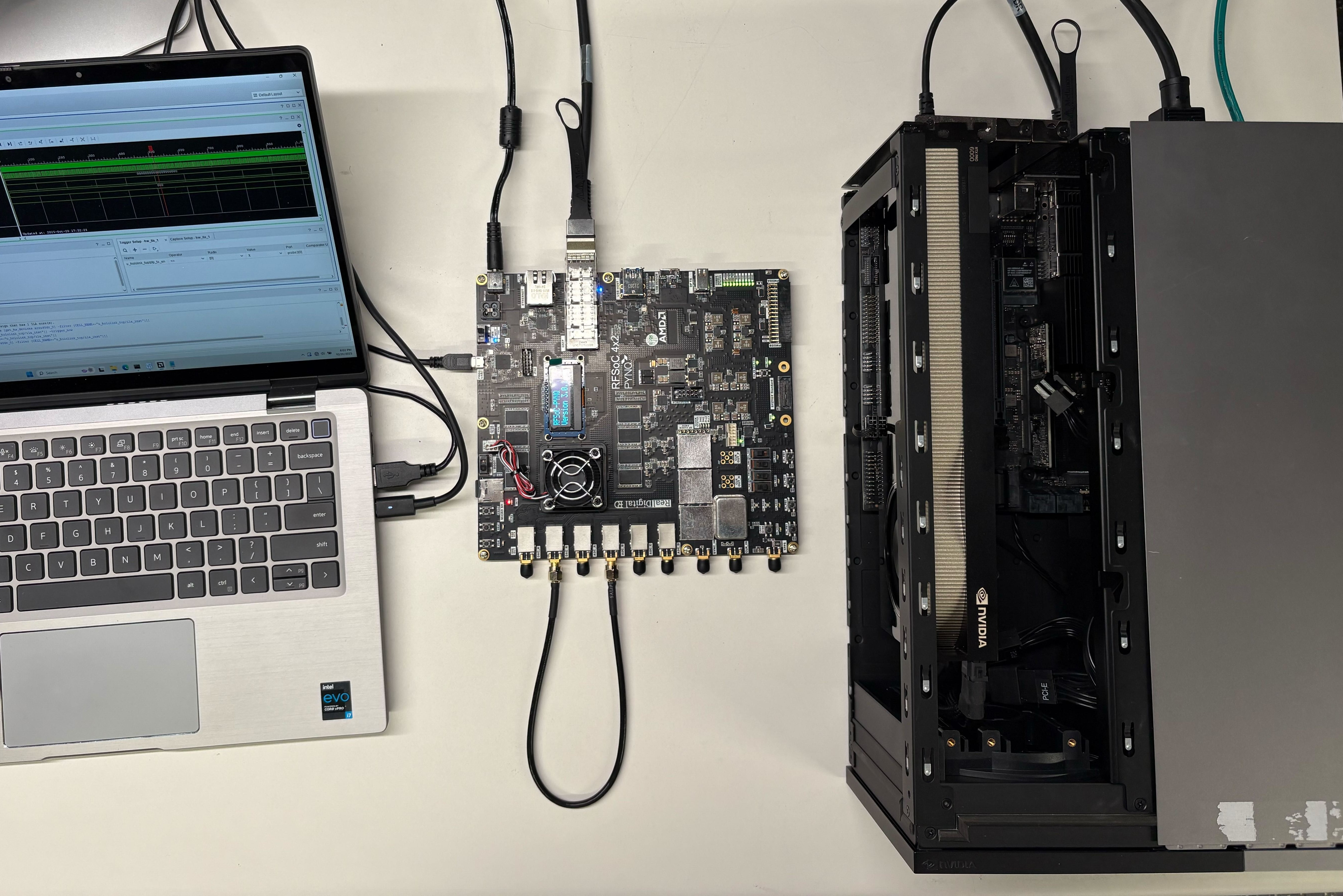}
    \caption{Proof of Concept setup}
    \label{fig:net_setup_photo}
\end{figure}

To show a Proof of Concept (PoC), we built on top of NVIDIA Holoscan Sensor Bridge (HSB)~\cite{hsb} ecosystem, which provides means to send data between an FPGA and NIC using the RDMA over Converged Ethernet (RoCE) protocol, as well handling the enumeration steps and the control signals. We made a setup with an ARM system hosting a NVIDIA RTX PRO 6000 Blackwell GPU and a NVIDIA ConnectX-7 Network Interface Card (NIC), as well as an AMD RFSoC FPGA, shown in Fig.~\ref{fig:net_setup_photo}. We also use a separate laptop to connect to the Integrated Logic Analyzer (ILA) within the FPGA to read out the results, not to have any impact on the latency measurements. This setup is using off the shelf components, and thanks to utilizing the the mature software stack of RoCE, it can be ported to other GPUs, such as NVIDIA GB300, and this GPU was only chosen for the PoC purposes. Clearly, the network card and GPU performance, and their connection over PCIe will have impacts on the latency.

To measure the latency, we designed a system where FPGA creates packets, host loops them back, and FPGA measures the time difference. To achieve high accuracy time measurements, a Precision Time Protocol (PTP) time stamp generator was used, to produce nanosecond level time values. The 96-bit value of PTP alongside a 16-bit value of packet number, along 18 bytes of 0 form the 32-byte payload for the RoCE packets. Upon arrival of the looped-back packets, the current time stamp and the timestamp in the packet, alongside its packet number, are sent to a laptop through the ILA. On the laptop, packets are checked to be consecutive, and per packet the time difference is measured.

\begin{figure}[ht]
    \centering
    \includegraphics[width=\linewidth]{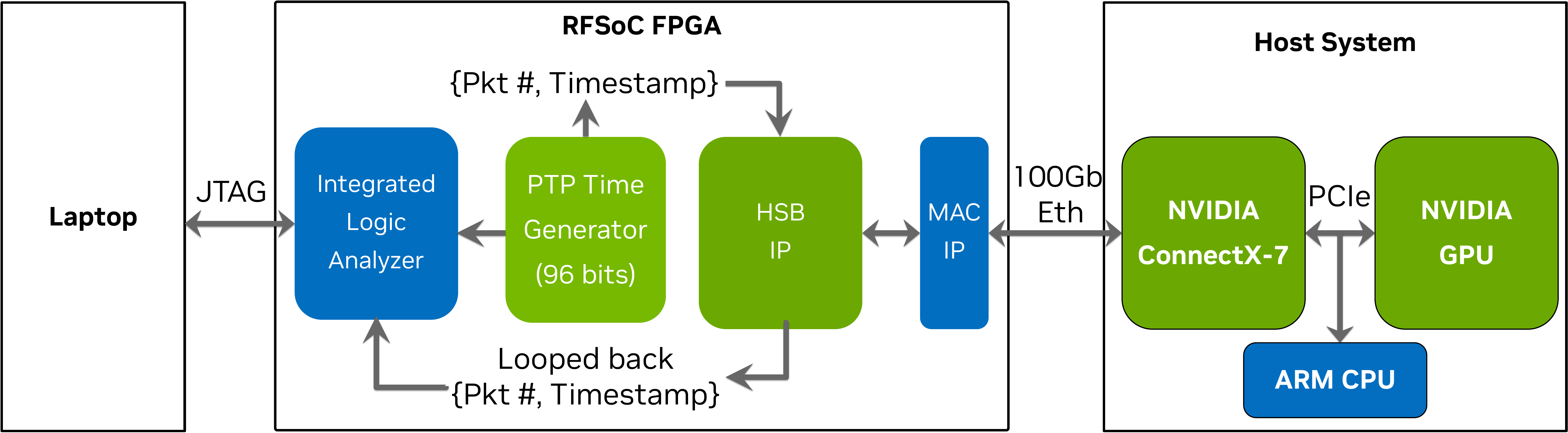}
    \caption{Proof of Concept flow}
    \label{fig:net_setup_diag}
\end{figure}

The main logical IP on the FPGA is the HSB IP, which is configured through our software and creates standard RoCE packets. This IP is available for several FPGA vendors, and follows the standard interfaces for the Ethernet MAC IPs. Moreover, on the receive side, it detaches the RoCE headers and delivers the payload to the FPGA. Additionally, this IP provides control signals for the FPGA, and we use them to set the time between packets, as well as start or stop the data stream from the FPGA. For a final design, the data stream can be changed from the packet number and time stamp to the proper data based on the application needs.

\begin{figure}[ht]
    \centering
    \begin{subfigure}[b]{0.49\linewidth}
        \centering
        \includegraphics[width=\linewidth]{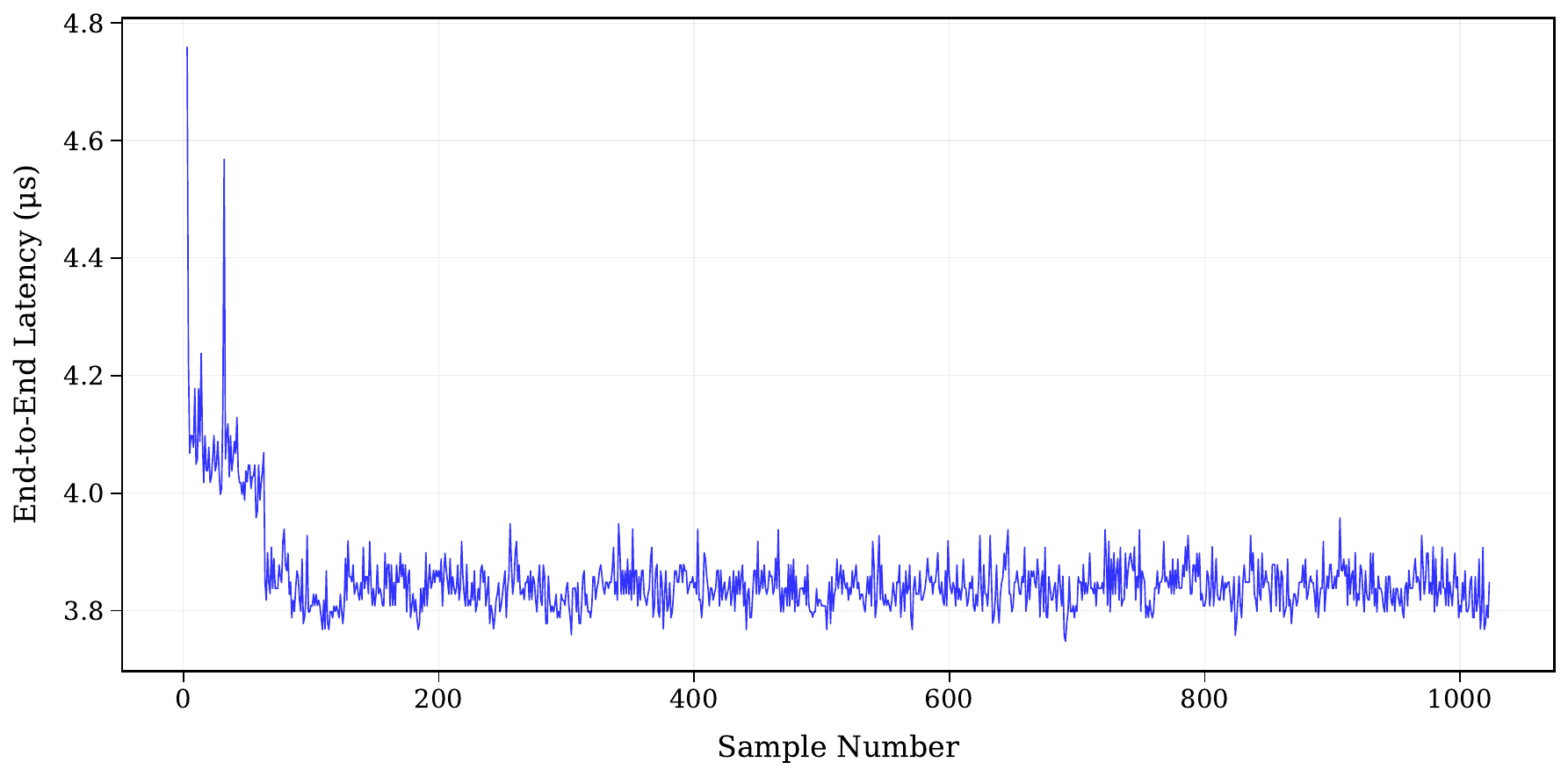}
        \caption{End-to-end latency over time}
        \label{fig:net_latency_start_raw}
    \end{subfigure}
    \hspace{0.02\linewidth}
    \begin{subfigure}[b]{0.4\linewidth}
        \centering
        \includegraphics[width=\linewidth]{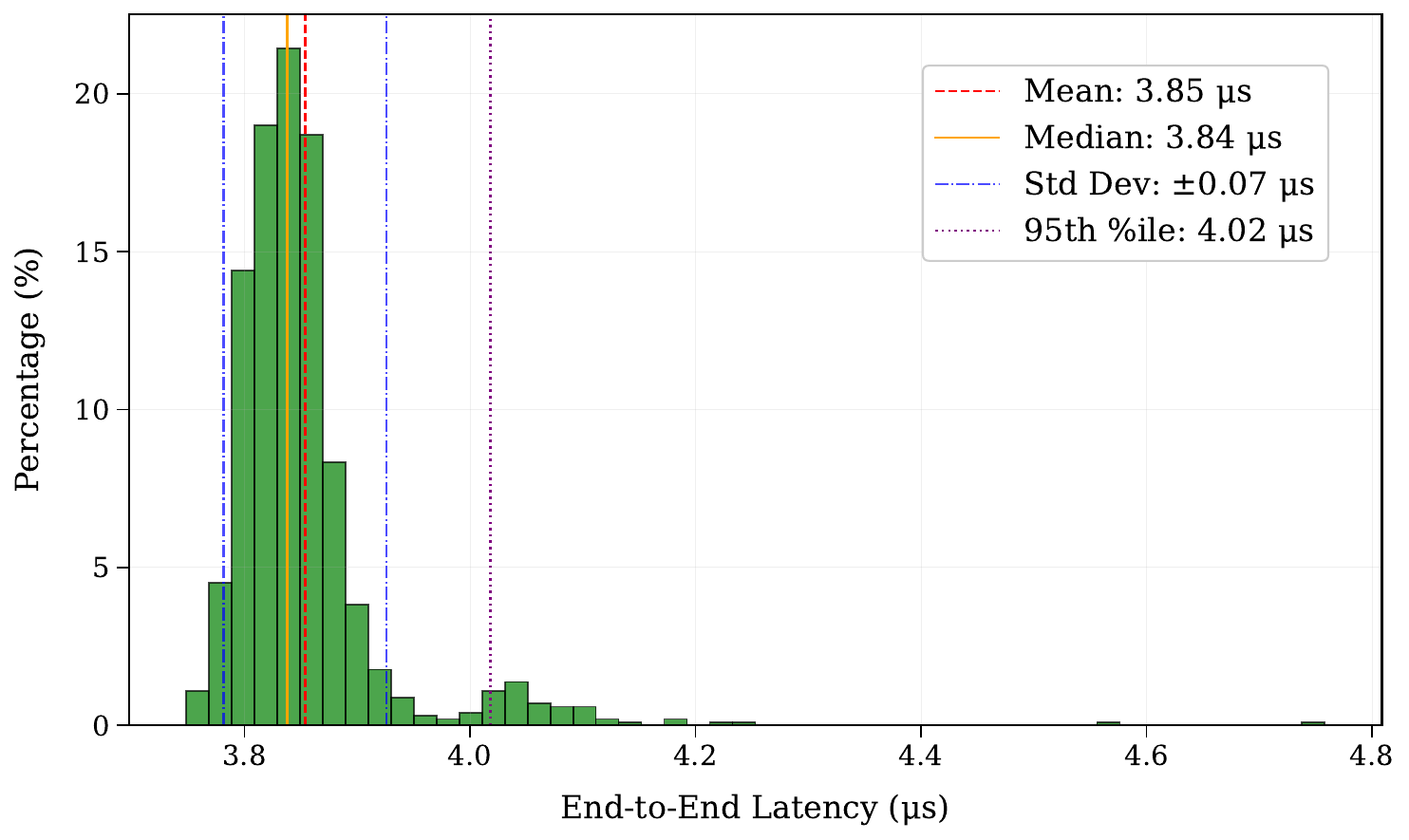}
        \caption{Histogram of end-to-end latency}
        \label{fig:net_latency_start_dist}
    \end{subfigure}
    \vspace{-7pt}
    \caption{Observing slight increased warm-up latency at the beginning of a run}
    \label{fig:net_latency_start}
\end{figure}

\begin{figure}[ht]
    \centering
    \begin{subfigure}[b]{0.49\linewidth}
        \centering
        \includegraphics[width=\linewidth]{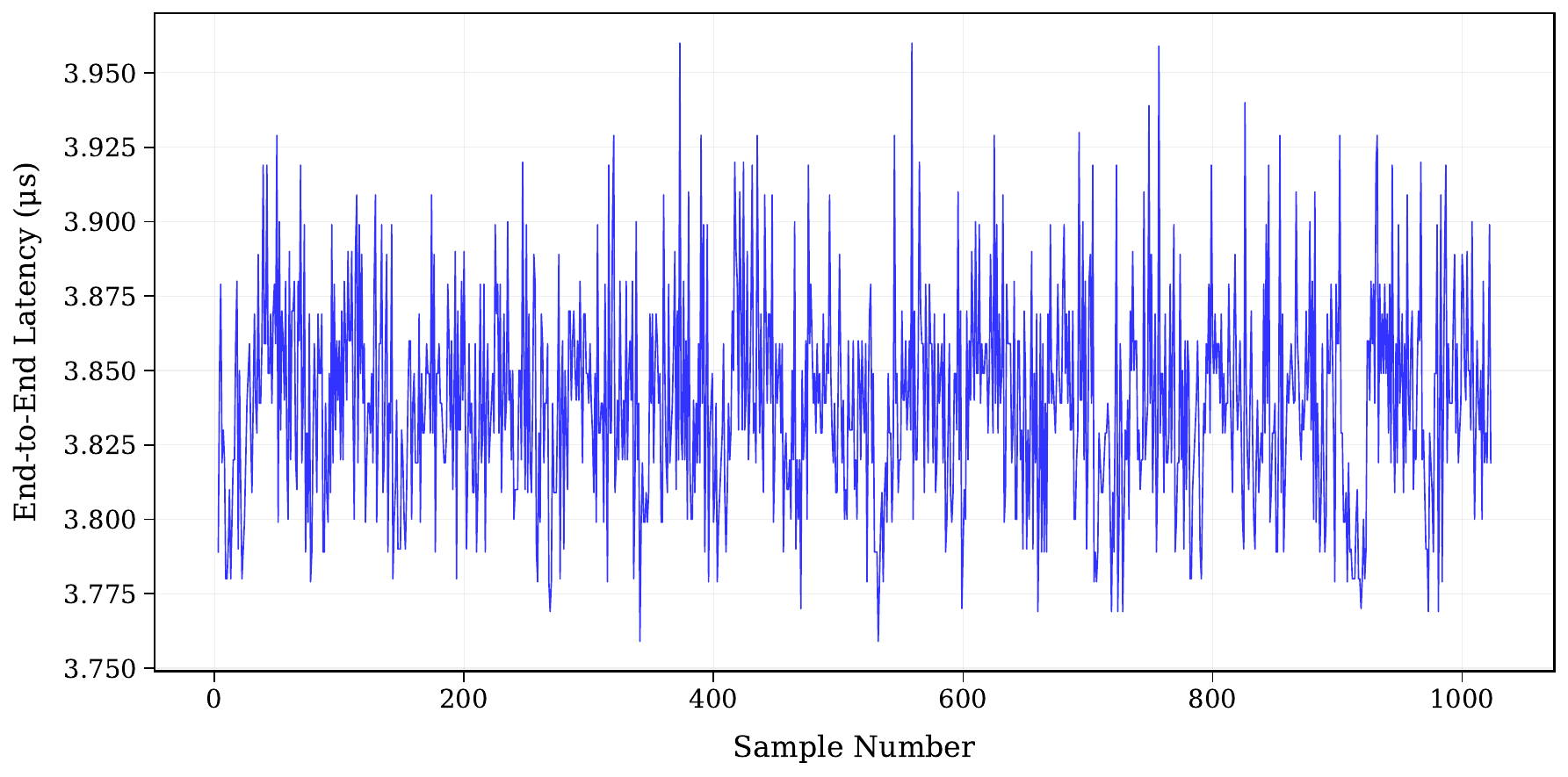}
        \caption{End-to-end latency over time}
        \label{fig:net_latency_mid_raw}
    \end{subfigure}
    \hspace{0.02\linewidth}
    \begin{subfigure}[b]{0.4\linewidth}
        \centering
        \includegraphics[width=\linewidth]{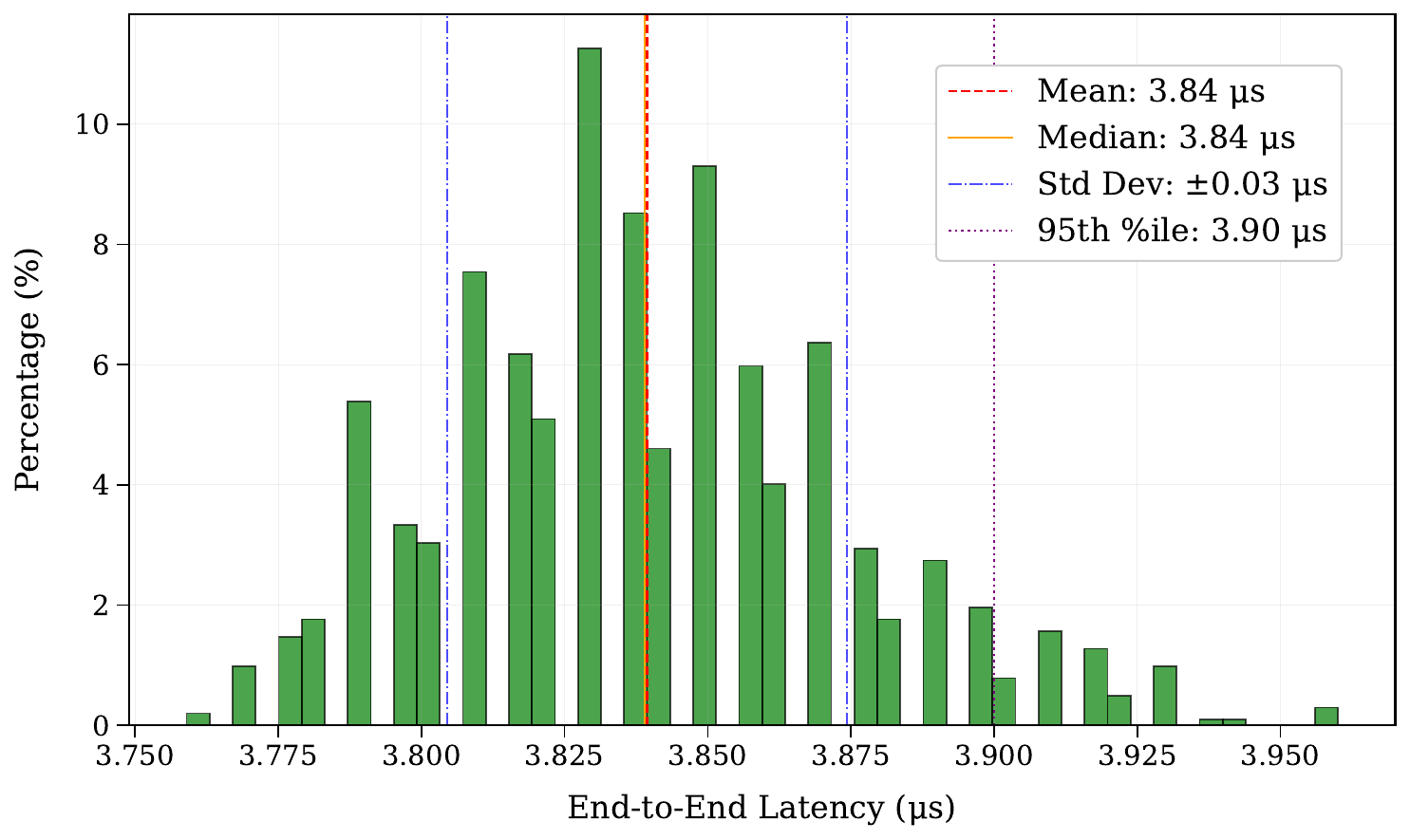}
        \caption{Histogram of end-to-end latency}
        \label{fig:net_latency_mid_dist}
    \end{subfigure}
    \vspace{-7pt}
    \caption{Steady state latency during the same run as Fig.~\ref{fig:net_latency_start}}
    \label{fig:net_latency_mid}
\end{figure}

Fig.~\ref{fig:net_latency_start_raw} shows the raw end-to-end latency captured by the ILA, and Fig.~\ref{fig:net_latency_start_dist} shows the distribution of this data. This is for the beginning of the run, as there is a need for a short warm-up period, which incurs higher latencies. This warm-up period is expected, due to factors such as some queue elements in the NIC being populated for the first time, or the cache allocation on NIC or GPU side. If desired, this can be removed through an initialization process with some dummy packets to cover this warm-up period. After that, the system becomes stable, and as shown in Fig.~\ref{fig:net_latency_mid_raw} and Fig.~\ref{fig:net_latency_mid_dist} for an example run, where we observe very little variance and end-to-end latencies below 4~$\mu$s. The mean and median latencies are measured at 3.839~$\mu$s, with a standard deviation of 35~ns and a sample maximum of 3.96~$\mu$s.

The loop-back software achieving this latency is comprised of two main parts, a CPU side that enumerates the FPGA and initializes the connection between the HSB IP and the host, as well as sending the extra desired configurations to the FPGA such as the data generation rate, and finally launches a persistent kernel on the GPU. The other part of the software is this persistent GPU kernel which takes over the processing on the host side, and waits for packet arrival, and loops them back. Note that benefiting from the RDMA technology, the data arrives at the GPU memory directly from the NIC, and benefiting from the GPUNetIO library, GPU directly sends the control commands and output packet to the NIC, without the host processor or its memory getting involved in any of these interactions.

\section{Programming Model}\label{scn:programming-model}

Programming \nvqlink systems is enabled by extensions to the open-source CUDA-Q platform that expose 
the heterogeneous quantum-classical architecture through a unified programming interface. 
This programming model extends CUDA's proven heterogeneous computing paradigm to encompass 
quantum processing units, enabling developers to orchestrate computation across CPUs, GPUs, 
and quantum control systems within a single application framework.

The cornerstone of this approach is the quantum kernel — a C++ function annotated with 
\literal{__qpu__} that seamlessly integrates quantum operations with classical control flow. 
These kernels are compiled by the \literal{nvq++} compiler into Multi-Level Intermediate Representation 
(MLIR) dialects: Quake for quantum instructions and CC for classical operations. This intermediate 
representation enables sophisticated optimization and provides the foundation for real-time 
quantum-classical communication through device callbacks.

\begin{definition}[Quantum Kernel]
A C++ function annotated with the \literal{__qpu__} attribute that encapsulates quantum operations 
within standard C++ control flow. Quantum kernels may contain quantum gates, measurements, and classical 
control logic, enabling the expression of complex quantum algorithms with familiar programming constructs.
\end{definition}

\begin{definition}[Device]
Within the CUDA-Q machine model, a device represents any computational resource within the logical QPU capable of 
executing library functions and kernels. Devices include traditional processors (CPUs, GPUs), quantum 
control systems (PPUs, Real-time Hosts), and composite systems (supercomputing clusters). Each device is 
assigned a unique identifier and can be composed hierarchically from other devices.
\end{definition}

\subsection{Heterogeneous Machine Model}

The CUDA-Q machine model for \nvqlink extends the familiar CUDA host-device paradigm across three 
computational domains, as illustrated in Table~\ref{tab:host-device}. This hierarchical abstraction enables 
programmers to reason about quantum-classical systems using established parallel programming concepts 
while accommodating the unique requirements of quantum control systems.

\begin{table}[ht]
\centering
\begin{tabularx}{\textwidth}{lXXX}
\textbf{Domain} & \textbf{Supercomputer} & \textbf{Real-time Host} & \textbf{QSC} \\
\hline\hline
\textbf{Programming Model} & CUDA-Q & CUDA & CUDA-Q\\
\textbf{Host} & Access Node CPU & Real-time Host CPU & Real-time Host CPU$^\dagger$\\
\textbf{Device} & Logical QPU & GPU & PPU or VPPU (emulator)\\
\textbf{Kernel Type} & 
    \makecell[l]{Quantum kernel:\\ \texttt{\_\_qpu\_\_}} & 
    \makecell[l]{CUDA kernel:\\ \texttt{\_\_global\_\_}} & 
    \makecell[l]{Device callback:\\ \texttt{cudaq::device\_call}} \\
\textbf{Compiler} & \texttt{nvq++} & \texttt{nvcc} & QSC-specific \\
\hline
\end{tabularx}
\caption{Programming model hierarchy spanning supercomputing, real-time control, and quantum control domains. $^\dagger$PPUs are addressable by the Real-time Host; QSCs may include dedicated control processors.}
\label{tab:host-device}
\end{table}

This three-tier architecture provides natural abstraction boundaries: the supercomputer manages
high-level algorithm orchestration, the Real-time Host coordinates real-time data management and error
correction, and the QSC executes fine-grained quantum control. Device unique identifiers
(UIDs) enable direct addressing across all tiers, while the Real-time Host maintains a
registry of available devices and their capabilities for efficient computation routing.

As indicated in~\autoref{tab:host-device}, the QSC domain supports both physical PPUs and Virtual Pulse Processing Units (VPPUs) as device types.
The VPPU provides a substitutable PPU emulator for offline development and testing (~\autoref{scn:vppu}).
From the programmer's perspective, quantum kernels compiled by \literal{nvq++} produce identical results when executed on VPPU or physical PPU, as both implement the same \literal{quantum_control_trait} interface (~\autoref{scn:devices}).
This substitutability enables developers to iterate on quantum programs and real-time protocols without requiring continuous physical hardware access, accelerating development workflows before deployment to production systems.

\subsection{Real-time Device Callbacks}

Central to enabling tight quantum-classical integration for CUDA-Q programmers is the \texttt{cudaq::device\_call} 
mechanism — a templated function intrinsic that enables quantum kernels to invoke computations on classical 
processing resources with automated data marshaling and minimal latency. This capability is essential 
for implementing real-time quantum error correction, adaptive algorithms, and measurement-based 
quantum computing protocols. Listing \ref{lst:qec-example} demonstrates what \texttt{device\_call} usage 
may look like to a CUDA-Q programmer.

\begin{lstlisting}[
    language=C++,
    caption={Real-time quantum error correction using device callbacks.},
    label={lst:qec-example},
    float=htbp
]
__qpu__ void adaptive_qec_kernel(cudaq::qvector<>& data_qubits, 
                                 cudaq::qvector<>& ancilla_qubits,
                                 int cycles) {
  for(int = 0; i < cycles; ++i){
    // Stabilizer circuits here
    ...
    // Execute syndrome extraction measurements
    auto syndrome = mz(ancilla_qubits);

    // Real-time streaming to dedicated GPU  
    cudaq::device_call(/*gpu_id=*/1, 
                     surface_code_enqueue, 
                     syndrome);
    // Repeat 
  }

  // Real-time decode on dedicated GPU  
  auto correction = cudaq::device_call(/*gpu_id=*/1, 
                                       surface_code_decode);
  
  // Apply corrections physically if desired (typically tracked in software)
  if (correction.x_errors.any()) 
    apply_pauli_x_corrections(data_qubits, correction.x_errors);
  if (correction.z_errors.any())
    apply_pauli_z_corrections(data_qubits, correction.z_errors);
}
\end{lstlisting}

The \texttt{device\_call} intrinsic supports multiple invocation patterns optimized for 
different computational scenarios. These templates signatures provide several key capabilities:

\begin{lstlisting}[
    language=C++,
    caption={Device callback template signatures.},
    label={lst:device-call-templates},
    float=htbp
]
// CPU function call on specified device
template <typename Callable, typename... Args>
auto device_call(std::size_t device_id, Callable&& func, Args&&... args) 
    -> decltype(func(std::forward<Args>(args)...));

// CPU function call on default device
template <typename Callable, typename... Args>
auto device_call(Callable&& func, Args&&... args) 
    -> decltype(func(std::forward<Args>(args)...));

// CUDA kernel launch with compile-time grid specification
template <std::size_t NumBlocks, std::size_t NumThreadsPerBlock,
          typename Kernel, typename... Args>
auto device_call(std::size_t device_id, Kernel&& kernel, Args&&... args)
    -> decltype(kernel(std::forward<Args>(args)...));
\end{lstlisting}

\begin{itemize}
\item \textbf{Device Selection}: Explicit targeting of computational resources by unique identifier
\item \textbf{Execution Flexibility}: Support for both CPU functions and GPU kernel launches  
\item \textbf{Automatic Marshaling}: Compiler-managed data movement between heterogeneous devices
\item \textbf{Type Safety}: Template-based interface preserving C++ type semantics and enabling compile-time error checking
\end{itemize}

Function signatures for device callbacks must conform to CUDA-Q type constraints. Arguments passed to 
device callbacks are considered immutable. Argument types must therefore be pass by value or based by constant reference (\texttt{const T\&}). 
Device functions must complete execution before quantum operations can resume, ensuring deterministic program semantics. 
Device callback library develpers are free to launch asynchronous threads within callback functions that return \texttt{void}. 

\subsection{Heterogeneous Memory Abstraction Layer}

Given the heterogeneous logical QPU architecture in \ref{fig:qpu}, the programming model requires a mechanism for reasoning 
about data allocated across inherent classical devices. Therefore, the programming model abstracts heterogeneous memory systems 
through the \texttt{cudaq::device\_ptr<T>} type, which encapsulates device-specific memory 
handles while maintaining type safety and automatic lifetime management. This abstraction 
enables transparent data movement optimized for the underlying hardware topology. 
To meet these requirements, we specify a \texttt{device\_ptr} type in Listing \ref{lst:device-pointer}.

\begin{lstlisting}[
  language=C++,
  caption={Device pointer abstraction for heterogeneous memory management.},
  label={lst:device-pointer},
  float=htbp
]
template <typename T>
struct device_ptr {
  std::size_t handle = std::numeric_limits<std::size_t>::max();
  std::size_t size = 0;
  std::size_t device_id = std::numeric_limits<std::size_t>::max();
  void* host_shadow = nullptr;
  
  device_ptr(T* host_data) : host_shadow(host_data) {
    handle = register_device_memory(host_data, sizeof(T));
    size = sizeof(T);
  }
};
\end{lstlisting}

The memory model supports both explicit control for performance-critical applications and implicit management 
for programmer productivity. Compiler implementations should analyze data usage patterns to optimize transfers, 
employing high-bandwidth RDMA for capable devices and efficient batching for network-attached resources. 
Memory coherency is maintained through a combination of compiler analysis and runtime synchronization, 
ensuring program correctness across the heterogeneous system.

\texttt{device\_ptr<T>} types provide a unique handle to allocated logical QPU device data, in a similar 
way that CUDA exposes device pointers for handling data on GPU device. These instances should 
be usable within CUDA-Q kernel code (i.e. quantum kernel signatures may contain \texttt{device\_ptr} types), 
enabling sophisticated quantum-classical workflows that mutate data on classical devices within a logical 
QPU via the \texttt{device\_call} intrinsic. 

This device data modeling also requires that NVQLink expose an API for creation, manipulation, and destruction 
of these device pointer types. In the same way that CUDA provides a device driver API 
(e.g. \texttt{cudaMalloc}, \texttt{cudaMemcpy}, \texttt{cudaFree}, \texttt{cuLaunchKernel}, etc.), 
we require a logical QPU device driver API, which we elaborate on in Section \ref{scn:device_driver}.

\subsection{Compilation Implications}

The \nvqlink concepts proposed here for the CUDA-Q programming model have direct ramifications on compiler implementations. 
Key to this is how one handles lowering the proposed \texttt{device\_call} intrinsic. It is expected that 
quantum operations naturally lower to pulse level representations, followed by lowering to subsequent operations 
that drive the dynamics of the quantum register via a distributed set of FPGAs or equivalent System on Chip (SoC) instances. 
Note we are intentionally generic in this last statement due to the potential spectrum quantum execution timing domains 
one may encounter. Pulse representations may lower directly to FPGA softcore-processor Instruction Set Architecture (ISA) code, or 
it may lower to standard CPU code that mediates pulse emission. Moreover, it is unclear how \texttt{device\_call} 
should lower to quantum control systems, and how modality-specific timing constraints should influence that 
code generation. Here we try to elucidate some of these finer points. 

For the task of quantum kernel compilation, we identify two end points of a system latency sensitivity spectrum 
that is dependent on the underlying system latency tolerance and associated control 
requirements - we classify this as high latency sensitivity vs low latency sensitivity.
Systems with high latency sensitivity are those that exhibit shorter qubit coherence times, thereby requiring 
lower real-time feedback latency. Systems with low latency sensitivity exhibit longer 
qubit coherence times, and can therefore tolerate slower real-time feedback. 
Each domain imposes different constraints on the communication 
patterns between the Real-time Host and the Quantum System Controller (QSC), fundamentally 
altering the kernel code compilation strategy.

\subsubsection{High Latency Sensitivity}
Quantum systems with stringent real-time requirements operate with extremely tight latency budgets that impose
the most restrictive timing constraints on the control system, therefore we enumerate the following
requirements for kernel compiler implementations:

\begin{itemize}
\item \textbf{No Real-time Host Mediation}: The control driver cannot mediate any data marshaling or function invocation 
during quantum operations due to latency constraints that would exceed decoherence timescales.

\item \textbf{Remote Direct Memory Access (DMA) Required}: All real-time classical processing must occur through Remote 
Direct Memory Access (RDMA) paths that bypass traditional network stacks and CPU involvement, enabling sub-microsecond data transfer.

\item \textbf{Pre-Compiled ISA Programs}: Complete ISA programs must be uploaded to 
FPGAs in advance and triggered atomically, with minimal interactive communication with the Real-time Host during execution. 
Control must be inherent to this pre-compiled ISA representation. We allow for dynamic instruction queuing, whereby 
FPGAs consuming ISA instructions pull the next instruction from a queue that is updated in real-time by the 
Real-time Host. The requirement then is that the instruction queue remains non-empty until program termination.

\end{itemize}

\begin{figure}[ht]
    \centering
    \includegraphics[width=0.9\linewidth]{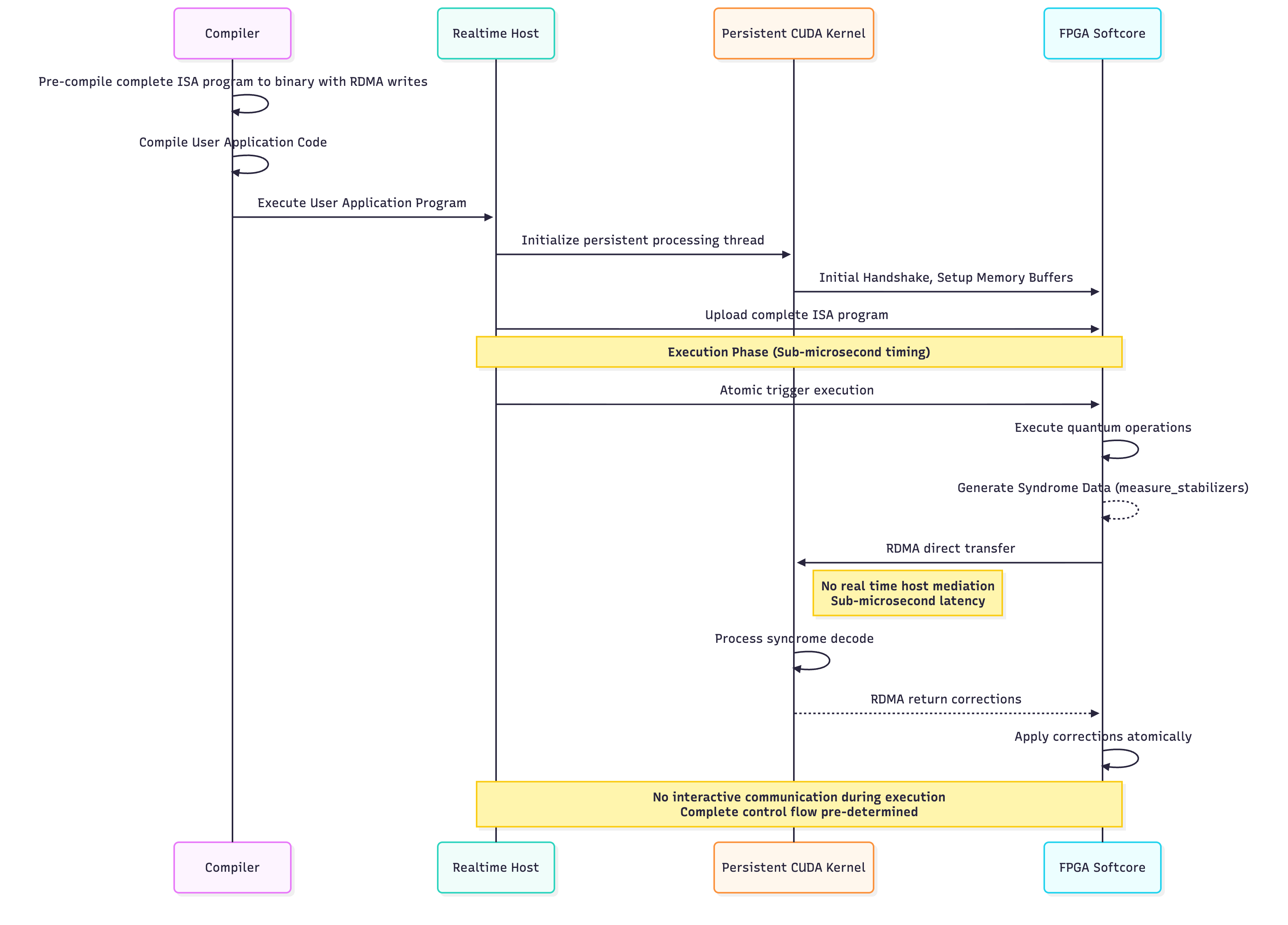}
    \caption{
        Sequence diagram describing the interaction between Real-time Host, CUDA Device, and FPGA control to enable 
        realtime data marshaling and classical device callback invocation for systems with 
        short coherence time requirements. Quantum kernels must 
        be compiled ahead-of-time to FPGA-control-specific ISA binary code. The Real-time Host must 
        initialize the total system - launching persistent CUDA threads that wait for FPGA-delivered data 
        for processing. 
    }
    \label{fig:fast_compiler_workflow}
% AJM Future Reference - here's the code to generate this sequence diagram
% sequenceDiagram
%     participant C as Compiler
%     participant RH as Real-time Host
%     participant PK as Persistent CUDA Kernel
%     participant FPGA as FPGA Softcore

%     %% Note over C,FPGA: Compilation Phase (Ahead-of-Time)
%     C->>C: Pre-compile complete ISA program to binary with RDMA writes
%     C->>C: Compile User Application Code
%     C->>RH: Execute User Application Program
%     RH->>PK: Initialize persistent processing thread
%     PK->>FPGA: Initial Handshake, Setup Memory Buffers 
%     RH->>FPGA: Upload complete ISA program
    
%     Note over RH,FPGA: Execution Phase (Sub-microsecond timing)
%     RH->>FPGA: Atomic trigger execution
%     FPGA->>FPGA: Execute quantum operations 
%     FPGA-->>FPGA: Generate Syndrome Data (measure_stabilizers)
%     FPGA->>PK: RDMA direct transfer 
%     Note right of PK: No real time host mediation<br/>Sub-microsecond latency
%     PK->>PK: Process syndrome decode
%     PK-->>FPGA: RDMA return corrections
%     FPGA->>FPGA: Apply corrections atomically
    
%     Note over RH,FPGA: No interactive communication during execution<br/>Complete control flow pre-determined
\end{figure}

The compilation model for high latency sensitivity systems must perform aggressive ahead-of-time optimization and leverage pre-initialized,
asynchronous data processing threads (e.g. CUDA persistent kernels) for real-time callback functionality. Figure \ref{fig:fast_compiler_workflow} demonstrates
this asynchronous workflow, and it is envisioned that compiler implementations will lower to this type of workflow for
latency-critical modalities.

\subsubsection{Low Latency Sensitivity}
Quantum systems with more relaxed timing requirements exhibit higher tolerance for processing delays,
enabling more flexible control architectures. Here we enumerate the requirements for kernel code lowering for
this timing domain:

\begin{itemize}
\item \textbf{Real-time Host Mediation Permitted}: The Real-time Host can mediate data marshaling and function invocation 
between quantum operations without violating coherence constraints.

\item \textbf{RDMA Beneficial but Optional}: While RDMA provides performance benefits, standard network communication 
paths remain viable for many applications.

\item \textbf{Runtime Operation Streaming}: The Real-time Host can stream operations at runtime.

\item \textbf{Interactive Execution Model}: Quantum and classical operations can be interleaved with bidirectional 
communication between the Real-time Host and QSC.
\end{itemize}

\begin{figure}[ht]
    \centering
    \includegraphics[width=0.9\linewidth]{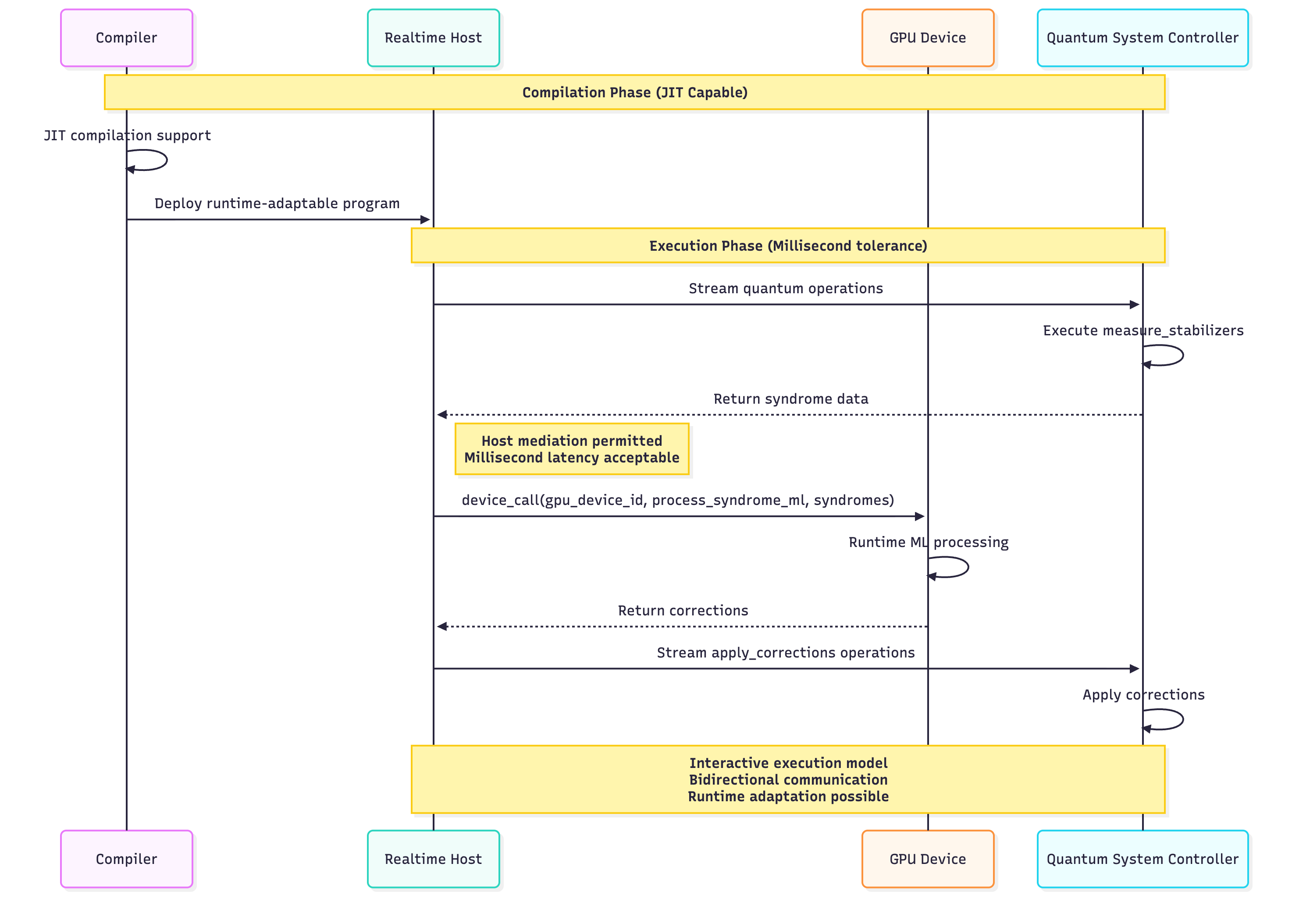}
    \caption{
        Sequence diagram describing the interaction between Real-time Host, CUDA Device, and FPGA control to enable 
        realtime data marshaling and classical device callback invocation for systems with 
        long coherence time requirements. Here there is more flexibility for the Real-time Host 
        to mediate interactions between FPGA control and GPU coprocessing. Quantum kernels can be 
        compiled ahead-of-time or just-in-time, and pulse level instructions can be dynamically generated and 
        emitted. Qubit measurement feedback can be mediated via the Real-time Host and marshaled manually to 
        available classical processing devices. 
    }
    \label{fig:slow_compiler_workflow}
    % AJM Future Reference - here's the code to generate this sequence diagram
    % sequenceDiagram
    % participant C as Compiler
    % participant RH as Real-time Host
    % participant GPU as GPU Device
    % participant QSC as Quantum System Controller

    % Note over C,QSC: Compilation Phase (JIT Capable)
    % C->>C: JIT compilation support
    % C->>RH: Deploy runtime-adaptable program
    
    % Note over RH,QSC: Execution Phase (Millisecond tolerance)
    % RH->>QSC: Stream quantum operations
    % QSC->>QSC: Execute measure_stabilizers
    % QSC-->>RH: Return syndrome data 
    % Note right of RH: Host mediation permitted<br/>Millisecond latency acceptable
    % RH->>GPU: device_call(gpu_device_id, process_syndrome_ml, syndromes)
    % GPU->>GPU: Runtime ML processing
    % GPU-->>RH: Return corrections
    % RH->>QSC: Stream apply_corrections operations
    % QSC->>QSC: Apply corrections
    
    % Note over RH,QSC: Interactive execution model<br/>Bidirectional communication<br/>Runtime adaptation possible
\end{figure}

The compilation model for low latency sensitivity systems clearly supports just-in-time (JIT) compilation,
optimization, and runtime adaptation. Figure \ref{fig:slow_compiler_workflow} demonstrates the envisioned
execution workflow for these systems. It is possible for compiler implementations to lower to
Real-time Host library calls that mediate and drive kernel execution, including real-time data marshaling and device callback
invocation.

\subsubsection{Lowering Workflows}
It will be helpful to elucidate what compiler lowering may look like in a typical implementation of an \nvqlink implementation.
\begin{figure}[htbp]
\begin{minipage}[t]{0.43\textwidth}
\begin{lstlisting}[
    style=cudaq,
]
// User code for a quantum kernel that 
// requires realtime device callback (add).

// Declare device function
int add(int, int);

// Define CUDA-Q quantum kernel
__qpu__ int maybe_increment(int i) {
  cudaq::qubit q;
  h(q);
  return cudaq::device_call(2, add, i, mz(q));
}
\end{lstlisting}
\end{minipage}
\hfill
\begin{minipage}[t]{0.52\textwidth}
\begin{lstlisting}[
    style=mlir
]
func.func @maybe_increment(%arg0: i32) -> i32 {
  %0 = quake.null_wire
  %1 = quake.h %0 : (!quake.wire) -> !quake.wire
  %measOut, %wires = quake.mz %1 : (!quake.wire) -> (!quake.measure, !quake.wire)
  %2 = quake.discriminate %measOut : (!quake.measure) -> i1
  %3 = cc.cast unsigned %2 : (i1) -> i32
  %4 = cc.device_call @add on 2 (%arg0, %3) : (i32, i32) -> i32
  return %4 : i32
}
\end{lstlisting}
\end{minipage}

\vspace{10pt}

\includegraphics[width=\textwidth]{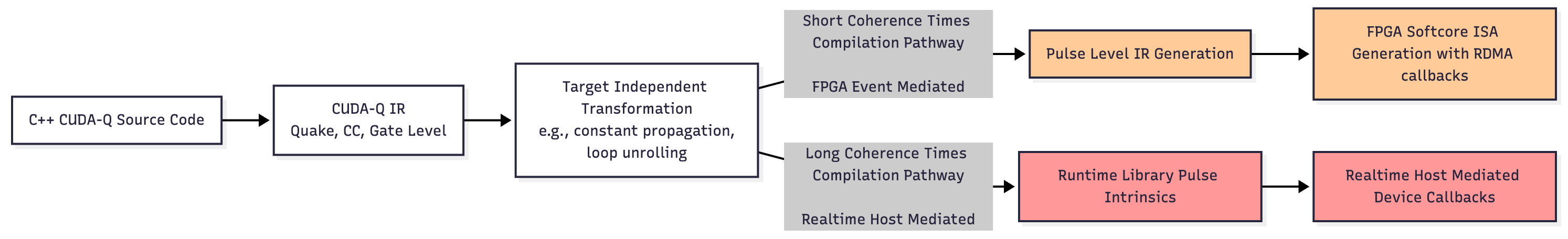}
\caption{(top left) Prototypical CUDA-Q quantum kernel code demonstrating quantum code 
interleaved with realtime classical processing on an \nvqlink device. This code executes 
on a Logical QPU and compiler code generation depends on the timing domain inherent to 
the targeted architecture. Compiler implementations, as a first step, should lower to a 
unified representation that is amenable for further target-specific lowering (top right).
Below these snippets (bottom), we see how compiler workflows should approach the proposed system 
timing domains.}
\label{fig:cudaq-mlir-lowering}
\end{figure}
Figure \ref{fig:cudaq-mlir-lowering} demonstrates the envisioned compiler code generation workflow for an \nvqlink architecture. 
Here one can see the quantum IR nodes (e.g. \texttt{quake.h}, etc.) as well as the critical \texttt{cc.device\_call} operation. It is expected that 
quantum operations are subsequently lowered to pulse level representations (e.g. Pulse MLIR Dialects) in tandem with dataflow into and out 
of this \texttt{device\_call} operation. The next phase of compilation is dictated by the inherent 
system timing domain exposed. Long coherence time domain systems may subsequently lower the code in Figure \ref{fig:cudaq-mlir-lowering} (top right) 
directly to a library of QSC-specific intrinsics on the Real-time Host. It is envisioned that the \texttt{device\_call} operation 
here will be lowered to a specific library function that takes as input callback arguments, the callback name, and a callback return value pointer and 
leverages the \nvqlink runtime API (see \ref{scn:runtime}) to generically affect callback invocation on a designated device within 
the Logical QPU. 

Latency-critical systems will require further lowering of pulse representations down to FPGA ISA code, with the understanding
that this is likely a distributed FPGA system, with specific physical qubits assigned to specific FPGAs. Hence,
the original unified kernel will need to be broken up into unique programs targeting specific FPGAs.
Moreover, implementations for this timing domain will require lowering of \texttt{device\_call} to
FPGA operations that trigger direct data marshaling from FPGA memory to GPU memory (RDMA). GPU device code
will need to be listening for data events asynchronously and apply requested callbacks once all data is received
from executing FPGA ISA code.

These compilation workflow concepts will span available \nvqlink architectural implementations due to the
enhancements we propose to the CUDA-Q programming model. Next, we turn our attention to necessary runtime support
for describing these compositional architectures and data types required for real-time classical coprocessing
during critical system timing windows.

\section{Runtime Architecture}\label{scn:runtime}

The \nvqlink runtime architecture promotes a high performance, zero-overhead abstraction model built on \emph{static polymorphism} and \emph{trait-based composition}.
These choices are made to serve the goals of minimizing real-time callback latency and maximizing configurability and extensibility of device types within the Logical QPU.

In this section, we fully specify required class types that enable expression of real-time data marshaling, data management, device 
callback processing for our envisioned logical QPU. Ultimately we propose interfaces for defining \emph{devices} and their unique behaviours 
or capabilities, as well as a general device driver API for the logical QPU. 

\subsection{Devices and their traits}\label{scn:devices}
To start, \nvqlink proposes a generic \literal{cudaq::device} type, shown in~\autoref{lst:cudaq-device}. This type serves as the 
fundamental abstraction for all processing resources within the Logical QPU.
Each device retains its own UID and a registry of its functions that can be invoked under a real-time callback during quantum kernel execution.
Each device receives its UID from an atomic counter during system initialization to ensure uniqueness.
Devices are intended to maintain a mapping from library locations to callable functions, enabling dynamic function resolution.
Devices expose \literal{connect()} and \literal{disconnect()} methods to enable one-time initialization and finalization of every device instance.

\begin{lstlisting}[
    language=C++,
    caption={Base class \literal{cudaq::device} for devices managed by the \nvqlink runtime.},
    label={lst:cudaq-device},
    float=htbp
]
template<typename Derived, typename... Traits>
class device {
private: 
  std::size_t device_id = 0;
public:  
  void connect();
  void disconnect(); 
  std::size_t get_id() const; 
};
\end{lstlisting}

The runtime expresses device functionality through a trait system that composes device-specific capabilities at compile time.
Each trait is a set of behaviors that can be implemented on an object and is expressed by a class whose members are all public methods.
We propose here a set of traits to be built into CUDA-Q, and this system supports straightforward extensibility to new traits defined by third parties.

We enumerate the following traits:

\begin{itemize}
\item \textbf{\texttt{explicit\_data\_marshalling\_trait}}: Provides realtime host mediated memory management with methods for allocation, deallocation, and 
  data transfer (see \ref{lst:explicit-data-marshaling-trait}). This trait enables real-time data processing for slow timing modalities. It exposes the familiar 
  API for data allocation, deallocation, and mutation. 
\item \textbf{\texttt{device\_callback\_trait}}: Enables realtime host mediated device function invocation (see \ref{lst:device-callback-trait}). This trait 
  enables real-time device callback capabilities for slow timing modalities. It exposes an API for applying a specified callback function (device-specific) 
  on provided function arguments represented as pre-allocated \texttt{device\_ptr} instances. It also allows for function result return via a separate 
  pre-allocated \texttt{device\_ptr} instance.
\item \textbf{\literal{quantum_control_trait}}: Specializes quantum control system functionality including program upload and trigger mechanisms.
  Devices that inherit this trait present a unified view of QCS devices for the realtime host (see \ref{lst:quantum-control-trait}). Programs are 
  represented as compiled kernel binaries. Execution is triggered with kernel arguments provided as \texttt{device\_ptr} instances. Kernel results 
  are returned via a pre-allocated \texttt{device\_ptr} instance. 
\item \textbf{\texttt{rdma\_trait}}: Supports high-performance, low-latency data transfer through remote direct memory access (see \ref{lst:rdma-trait}). This 
  trait is designed for fast architecture modalities where real-time host mediation of data marshaling and callback invocation is not possible. It 
  exposes a minimal interface allowing one to extract memory buffer data for one-time initialization and handshake mechanisms with designated \literal{quantum_control_trait} 
  device instances. Concrete \texttt{rdma\_trait} types are required to perform all pertinent RDMA connection tasks within the sub-type specific \texttt{connect} method implementation.
\end{itemize}

\begin{lstlisting}[
    language=C++,
    caption={Trait supporting data movement controlled by the Real-time Host.},
    label={lst:explicit-data-marshaling-trait},
    float=htbp
]
template <typename Derived>
class explicit_data_marshaling_trait {
public:
  void *resolve_pointer(device_ptr &devPtr);

  device_ptr malloc(std::size_t size) const;

  template <typename... Sizes,
            std::enable_if_t<(std::conjunction_v<std::is_integral<Sizes>...>),
                             int> = 0>
  auto malloc(Sizes... szs) {
    return std::make_tuple(static_cast<Derived *>(this)->malloc(szs)...);
  }

  void free(device_ptr &d);

  template <typename... Ptrs,
            typename = std::enable_if_t<
                (std::conjunction_v<std::is_same<
                     std::remove_cv_t<std::remove_reference_t<Ptrs>>,
                     device_ptr>...>)>>
  void free(Ptrs &&...d) {
    (free(d), ...);
  }

  void send(device_ptr &dest, const void *src);
  void recv(void *dest, const device_ptr &src);
};
\end{lstlisting}

\begin{lstlisting}[
    language=C++,
    caption={Trait supporting real-time device callback functions.},
    label={lst:device-callback-trait},
    float=htbp
]
template <typename Derived>
class device_callback_trait {
public:
  void launch_callback(const std::string &callbackName,
                       const std::vector<device_ptr> &args);
  void launch_callback(const std::string &callbackName, device_ptr &result,
                       const std::vector<device_ptr> &args);
};
\end{lstlisting}

\begin{lstlisting}[
    language=C++,
    caption={Trait supporting program binary upload and execution on QCS devices.},
    label={lst:quantum-control-trait},
    float=htbp
]
template <typename Derived>
class quantum_control_trait {
public:
  void upload_program(const std::vector<std::byte> program_data);
  void trigger(device_ptr &result, const std::vector<device_ptr> &args, const cudaqx::heterogeneous_map& config);
};
\end{lstlisting}

\begin{lstlisting}[
    language=C++,
    caption={Trait supporting data movement by remote direct memory access (RDMA).},
    label={lst:rdma-trait},
    float=htbp
]
template <typename Derived, typename RDMADataType>
class rdma_trait {
public:
  RDMADataType& get_rdma_connection_data() const;
};
\end{lstlisting}

This device architecture is designed to support various timing modalities and real-time data processing capabilities. 
Instantiation of concrete device types allows users to define unique logical QPU architectures programmatically. 
Their use alongside architecture-specific quantum kernel compilation allows the development of real-time 
quantum-classical workflows that leverage GPU coprocessing natively. 

\subsection{Compiled Quantum Kernels}\label{scn:compilation}

The \nvqlink architecture considers general quantum kernel compilation as the means for mapping user-intent 
to executable instructions on pulse processors. We do not limit ourselves to static circuit compilation, but 
instead consider fully parameterized functional representations that leverage general classical control 
flow. Moreover, we allow for control flow that is non-deterministic (e.g. dictated by quantum measurement results). 
We take a CUDA-Q kernel centric approach, but seek to enable 3rd party compilation and binary loading as well, as 
long as 3rd party kernels support our proposed \texttt{device\_call} capability. 

We promote an ahead-of-time compilation strategy and rely on standard code linking techniques to construct 
hybrid quantum-classical executables and libraries. As such, the \nvqlink architecture proposes an extension 
point for the library that enables compiled kernel generation. We represent compiled kernel code as in \ref{lst:compile_kernel}. 

\begin{lstlisting}[
    language=C++,
    caption={The \nvqlink representation for compiled quantum kernels.},
    label={lst:compile_kernel},
    float=htbp
]
struct qcontrol_program {
  std::vector<std::byte> binary;
  std::size_t qdevice_id;
};
class compiled_kernel {
protected:
  std::string kernel_name;
  std::vector<qcontrol_program> control_programs;
public:
  const std::string &name() const;
  const std::vector<qcontrol_program> &get_programs() const;
};
\end{lstlisting}

\begin{lstlisting}[
    language=C++,
    caption={The \nvqlink representation for architecture-specific compilers.},
    label={lst:compiler_class},
    float=htbp
]
struct qcontrol_program {
  std::vector<std::byte> binary;
  std::size_t qdevice_id;
};
class compiled_kernel {
protected:
  std::string kernel_name;
  std::vector<qcontrol_program> control_programs;
public:
  const std::string &name() const;
  const std::vector<qcontrol_program> &get_programs() const;
};
\end{lstlisting}

We represent the compiled quantum kernel as a singular unit, but composed of potentially many individual binary programs (the \texttt{qcontrol\_program}). 
Logical QPUs are expected to be composed of many \literal{quantum_control_trait} devices (e.g. many pulse processors) and this 
design allows for the expression of binary programs for each constituent control device. Each program 
tracks that it is intended to run on. Quantum kernels in CUDA-Q are assigned a kernel name via its function name and signature. We 
retain this information and tag it to the \texttt{compiled\_kernel} instance. 

To generate compiled kernel instances, \nvqlink proposes a \texttt{compiler} interface that is meant for architecture-specific compilation extensibility. 
The class is defined as defined in \ref{lst:compiler_class}

This API enables architecture-specific pre-compiled code loading (e.g. from object files) as well 
as JIT code compilation. Each capability takes as input the architecture-defined \literal{quantum_control_trait} 
devices so that the compiled kernel is fully architecture-aware. 

\subsection{Holistic Kernel Execution}\label{scn:executor}
The \nvqlink architecture promotes a hardware-aware capability for triggering kernel 
launches on available and specified \literal{quantum_control_trait} devices. 

\begin{lstlisting}[
    language=C++,
    caption={The \nvqlink representation for architecture-specific kernel executors.},
    label={lst:executor_class},
    float=htbp
]
template <typename Derived>
class executor {
public:
  void
  trigger_execution(std::unordered_map<std::size_t, any_device> &qcs_devices,
                    device_ptr &res, const std::vector<device_ptr> &args);
};
\end{lstlisting}

The intention for kernel launching is that the \nvqlink implementation will 
upload all required programs to specified \literal{quantum_control_trait} devices. Once 
all binary programs have been uploaded, an architecture-specific \texttt{executor} 
will be used to trigger the synchronous execution of all binary pulse programs. 

The \texttt{executor} provides a singular view on the collective pulse processing units, effectively 
modeling a single virtual pulse processor. This type may be leveraged for more than just hardware-specific 
program triggering. As an effective adaptor on quantum code execution, one could leverage this extension point 
to provide real-time instruction streaming and scheduling. 

\subsection{Logical QPU Driver API}\label{scn:device_driver}

To this point, we have proposed an architecture for constructing Logical QPU instances that enable 
real-time (within coherence times) data marshaling and processing. For envisioned advanced use cases, 
it will prove beneficial to define a library API on top of the \nvqlink proposed types enabling 
efficient Logical QPU data management, mutation, and kernel and device callback invocation. To this end, 
we propose an \nvqlink Driver API exposing a familiar interface for coprocessor data allocation, deallocation, 
mutation, and kernel loading and launching. 

\begin{lstlisting}[
    language=C++,
    caption={The \nvqlink Logical QPU Driver API.},
    label={lst:driver_api},
    float=htbp
]
template <typename... DeviceTypes>
void initialize(DeviceTypes &&...in_devices);
void shutdown();

std::size_t get_num_qcs_devices();
std::size_t get_num_classical_devices();

device_ptr malloc(std::size_t size);
device_ptr malloc(std::size_t size, std::size_t devId);
template <typename T>
device_ptr malloc(std::optional<std::size_t> device_it = std::nullopt);

void free(device_ptr &d);
template <
    typename... Ptrs,
    typename = std::enable_if_t<
        (std::conjunction_v<std::is_same<
             std::remove_cv_t<std::remove_reference_t<Ptrs>>, device_ptr>...>)>>
void free(Ptrs &&...d) {
  (free(d), ...);
}

void memcpy_to_qpu(device_ptr &arg, const void *src);
void memcpy_from_qpu(void *dest, const device_ptr &src);
template <typename T>
T memcpy_from_qpu(const device_ptr &src) ;

handle load_kernel(const std::string &location,
                   const std::string &kernel_name);
handle load_kernel_from_code(const std::string &code,
                             const std::string &kernel_name);

void launch_kernel(handle kernelHandle, device_ptr &result,
                   const std::vector<device_ptr> &args);
void launch_kernel(handle kernelHandle, const std::vector<device_ptr> &args);
template <typename Ret, typename... Args>
Ret launch_kernel(handle kernelHandle, Args &&...args);
\end{lstlisting}

\nvqlink proposes the function API described in \ref{lst:driver_api}. 
The library specification starts with general initialization and shutdown functions. Initialization 
takes as input the user-specified concrete devices that compose the Logical QPU. 

The library proposes the familiar user-facing set of data allocation and deallocation functions (\texttt{malloc} and \texttt{free}). 
These functions allow one to specify a target device with the Logical QPU where the data resides. If one does not specify 
a concrete device ID, the real-time host memory space is targetd. These functions return and 
operate on the aforementioned \texttt{device\_ptr} instances. The library promotes a device-specific data mutation 
API (\texttt{memcpy}) allowing one to move data to and from the Logical QPU. 

The API exposes a mechanism for loading and launching quantum kernels. The library should be configurable in this regard, 
allowing one to specify architecture-specific compiler and executor instances. Loading a pre-compiled kernel from file, 
or JIT compiling from source string, delegate to the specified compiler implementation and return a 
handle to the compiled kernel. This handle is used for subsequent calls to \texttt{launch\_kernel}. Kernel 
launching takes as input the kernel arguments, pre-allocated on the real-time host via this API. For 
kernels that return a value, one can supply an appropriately sized \texttt{device\_ptr} to \texttt{launch\_kernel}. 

\section{Preliminary Specification}\label{scn:spec}

We present here a prospective set of requirements here in an effort to telegraph future work toward a hardened specification and inspire readers to provide feedback and input on that specification.

\begin{enumerate}
\item The firmware definition of the PPU MUST remain reprogrammable by the QSC builder in the field.
\item The \nvqlink system MUST support an operating mode in which PPU instructions are opaque to the Real-time Host.
    In particular, the operator must have the option of encrypting these instructions without losing functionality in the Real-time Domain.
\item The Network Interface MUST be guaranteed to interoperate with the Real-time Host and Interconnect, and the performance characteristics of the network SHOULD be publicly documented and readily measurable by end users.
\item The Real-time Host MUST be programmable by \literal{C++}, CUDA, and CUDA-Q.
\item The Real-time Host MAY include other processing resources, including CPU-GPU systems-on-chip (SoCs), FPGAs, or ASICs.
\item The QSC provider MAY provide a Virtual Pulse Processing Unit (VPPU) that emulates its PPU instruction set for offline development and testing (~\autoref{scn:vppu}).
\item Round trip data latency from QSC to Real-time Host and back MUST be measurable by a CUDA-Q library function, and the results of this function invocation MUST return latency of every measured sample over its set of input (to be determined).
\end{enumerate}

\section{QPU Level Workloads}\label{scn:qpu-workloads}

This section translates \nvqlink abstractions into concrete QPU-level workloads.
We first work through a minimal but representative QEC subroutine, $T$-state teleportation, to illustrate kernel-embedded device callbacks and decoder interaction (\autoref{scn:t-gate-teleportation}).
We then analyze scalable fault-tolerant execution under lattice surgery, connecting decoder throughput and reaction time to parallel-window strategies and GPU batch execution (\autoref{scn:scalable-ftq-programs}).
Finally, we describe how the availability of HPC with tight coupling can benefit calibration and QCVV workloads with fast, parameterized control in (\autoref{scn:online-calibration} and \autoref{scn:benchmarking}).

\subsection{Example: $T$ gate teleportation in CUDA-Q}\label{scn:t-gate-teleportation}

Experimental demonstrations of fault-tolerant quantum computing subroutines are an active research topic in quantum computing and an essential step toward practical quantum computation.
To move beyond experimentation to production, developers need access to new primitives in quantum-capable programming languages, including real-time processing of syndrome data and low-latency data exchange between FPGAs and GPUs.

A canonical subroutine that depends on these real-time capabilities is magic state distillation and teleportation.
Once a magic state is prepared on a logical ancilla qubit through magic state distillation, cultivation, or similar methods, it can be applied to a target qubit via teleportation~\cite{Bravyi2005distillandteleport, gidney2024magicstatecultivationgrowing}.
The logical measurement outcome of the ancilla qubit determines whether a conditional $S$ gate must be applied to complete the teleportation protocol.
This measurement step is where the decoder may block execution, since the syndrome history from both the target and ancilla qubits is analyzed to predict fault locations and determine whether the logical observable was flipped.

The following example in~\autoref{lst:t-example-main} illustrates the key components involved in calling out to a decoder from CUDA-Q to perform $T$ gate teleportation.

The decoder is first initialized and waits for data, while the CUDA-Q kernel is then launched, sending calls to the decoder kernel during execution.
Much of the complexity lies in configuring the decoder server to understand what data to expect, how to convert raw stabilizer measurements into detector events, and which logical observables need to be decoded.
\begin{lstlisting}[
    language=C++,
    caption={CUDA-Q C++ main for a $T$ gate teleportation example.},
    label={lst:t-example-main},
    float=htbp
]
// Client/Server decoder example
int main(){
  // Now configure the decoders
  // Setting up the decoder properly requires information about the structure of the quantum kernel
  auto config = config_from_kernel(example_kernel, nData, nAncx, nAncz, x_schedule, z_schedule);
  cudaq::qec::decoding::config::configure_decoders(config);

  // Set up experiment parameters
  // ...
  // Here an example kernel is parametrized on physical qubits in a logical qubit.

  auto run_result = cudaq::run(numShots, example_kernel, nData, nAncx, nAncz, x_schedule, z_schedule);

  parse_logical_results(run_result);
  cudaq::qec::decoding::config::finalize_decoders();
}
\end{lstlisting}

\autoref{lst:t-example-entrypoint} shows the example kernel which has the main logic for the teleportation routine.
The $T$ gate production itself can be any routine, but many have a repeat-until-success component.
We do not implement \literal{t_distill_attempt}, but show and example which sits under a while loop until it returns that the protocol has been successful.
In this case, simple flag checking is done under the hood, but more generally this can require calls to a decoder as well.

\begin{lstlisting}[
    language=C++,
    caption={CUDA-Q kernel body, setting up and performing the teleportation.},
    label={lst:t-example-entrypoint},
    float=htbp
]
// Client/Server style decoding
void __qpu__ example_kernel(int nData, int nAncx, int nAncz,
           const std::vector<std::size_t> &x_schedule,
           const std::vector<std::size_t> &z_schedule,
           int nRounds){
  // Allocate logical qubit 0
  cudaq::qvector data_q(nData);
  cudaq::qvector anc_x(nAncx);
  cudaq::qvector anc_z(nAncz);

  // lQ0 will distill t state
  cudaq::qec::patch logicalQ0(nData, nAncx, nAncz);

  // Example repeat-until-success distillation protocol
  bool success = false;
  while(success == false){
    success = t_distill_attempt(logicalQ0);
  }

  // lQ1 will be target of teleportation
  cudaq::qec::patch logicalQ1(nData, nAncx, nAncz);

  // 1 device in this example.
  int deviceID = 0;
  // 1 decoder in this example.
  int decoderID = 0;
  // Certain systems may require data tagging
  int tagID = 0;

  // Run stabilizer rounds to initialize into lQ1 into logical |0> state,
  // and preserve lQ0
  for(int r = 0; r < nRounds; ++r){
    // Specify how syndrome bits are sent to decoders
    stabilizer_round(deviceID, tagID, logicalQ0, x_schedule, z_schedule);
    tagID += 1;
    stabilizer_round(deviceID, tagID, logicalQ1, x_schedule, z_schedule);
    tagID += 1;
  }

  // transversal CX
  cx(logicalQ1, logicalQ0);

  // Additional QEC cycles
  for(int r = 0; r < nRounds; ++r){
    // Specify how syndrome bits are sent to decoders
    stabilizer_round(deviceID, tagID, logicalQ0, x_schedule, z_schedule);
    tagID += 1;
    stabilizer_round(deviceID, tagID, logicalQ1, x_schedule, z_schedule);
    tagID += 1;
  }

  auto data_readout0 = mz(logicalQ0.data);
  auto readout_packed = cudaq::to_integer(data_readout0);

  cudaq::device_call(custom_library::enqueue_syndromes_ui64,
      deviceID, readout_packed, decoder_id, tagID);

  bool correction = cudaq::device_call(custom_library::get_correction,
                        deviceID, readout_packed, decoder_id);

  // Conditional s gate to correct t state teleportation
  if(correction){
    s(logicalQ1);
  }

  // Depending on the system, may need additional rounds of QEC while waiting on decoder.
  // Here we simply readout.
  auto data_readout1 = mz(logicalQ1.data);
  return data_readout1;
}

\end{lstlisting}

The bulk of gates executed in this example come from the rounds of stabilizer measurements.
This is where most of the syndrome data is passed to the decoder.

\begin{lstlisting}[
    language=C++,
    caption={CUDA-Q sub-kernel performing stabilizer rounds with data transfers to a decoder.},
    label={lst:t-example-stabs},
    float=htbp
]
// X stabilizers and Z stabilizers specified via cnot schedules
__qpu__ std::vector<cudaq::measure_result>
stabilizer_round(int deviceID, int tagID, patch logicalQubit, const std::vector<std::size_t> &x_schedule,
           const std::vector<std::size_t> &z_schedule) {
  for (std::size_t i = 0; i < logicalQubit.ancx.size(); i++)
    reset(logicalQubit.ancx[i]);
  for (std::size_t i = 0; i < logicalQubit.ancz.size(); i++)
    reset(logicalQubit.ancz[i]);

  // x_stabilizer circuit
  h(logicalQubit.ancx);
  for (std::size_t xi = 0; xi < logicalQubit.ancx.size(); ++xi)
    for (std::size_t di = 0; di < logicalQubit.data.size(); ++di)
      if (x_schedule[xi * logicalQubit.data.size() + di] == 1)
        cudaq::x<cudaq::ctrl>(logicalQubit.ancx[xi], logicalQubit.data[di]);
  h(logicalQubit.ancx);

  // z_stabilizer circuit
  for (size_t zi = 0; zi < logicalQubit.ancz.size(); ++zi)
    for (size_t di = 0; di < logicalQubit.data.size(); ++di)
      if (z_schedule[zi * logicalQubit.data.size() + di] == 1)
        cudaq::x<cudaq::ctrl>(logicalQubit.data[di], logicalQubit.ancz[zi]);

  // Decoder convention dependent:
  // S = (S_X, S_Z), (x flip syndromes, z flip syndrones).
  // x flips are trigger z-stabilizers (ancz)
  // z flips are trigger x-stabilizers (ancx)
  auto syndrome_bits = mz(logicalQubit.ancz, logicalQubit.ancx);

  // Can use CUDA-Q wrapped call for device portability
  // int decoderID = deviceID;
  // cudaq::qec::decoding::enqueue_syndromes(decoderID, syndrome_bits);

  // Or call custom library implementation
  auto syndrome_packed = cudaq::to_integer(syndrome);
  cudaq::device_call(custom_library::enqueue_syndromes_ui64,
                      deviceID, syndrome_packed, decoder_id, tagID);
  return syndrome_bits;
}
\end{lstlisting}
In this example, the $T$ gate is teleported onto a freshly initialized logical qubit, but the routine also demonstrates how teleportation can be applied to a logical qubit in a general quantum state.
In such cases, the qubit persists beyond this step and participates in subsequent Clifford and non-Clifford gate operations, rather than being immediately measured as shown here.

What is not shown in the example are the explicit \literal{cudaq::device_call} statements.
This omission highlights an important aspect of the approach: different hardware platforms have varying requirements for classical interactions, and a flexible interface for invoking classical resources from within a quantum kernel is therefore essential.
The \literal{cudaq::device_call} mechanism enables such flexibility.
In general, CUDA-Q provides high-level APIs designed to support multiple qubit technologies, while still allowing a modular, library-centric framework in which hardware developers can integrate their own solutions through \literal{cudaq::device_call}.

In latency-sensitive regimes, the quantum kernels may resemble the example shown above, where only the minimum dynamism required—such as repeat-until-success logic and conditional S gates—is enabled.
On hardware with greater tolerance for latency, the design can permit more frequent device calls.
For instance, the decoder could be reconfigured dynamically based on intermediate computation results.
Just-in-time compilation for lattice surgery routines is another promising direction that could lead to more efficient routing of logical two-qubit operations.

\subsection{Scalable fault tolerant quantum programs}\label{scn:scalable-ftq-programs}

\begin{figure}[ht]
    \centering
    \begin{subfigure}[b]{0.7\linewidth}
        \centering
        \includegraphics[width=\linewidth]{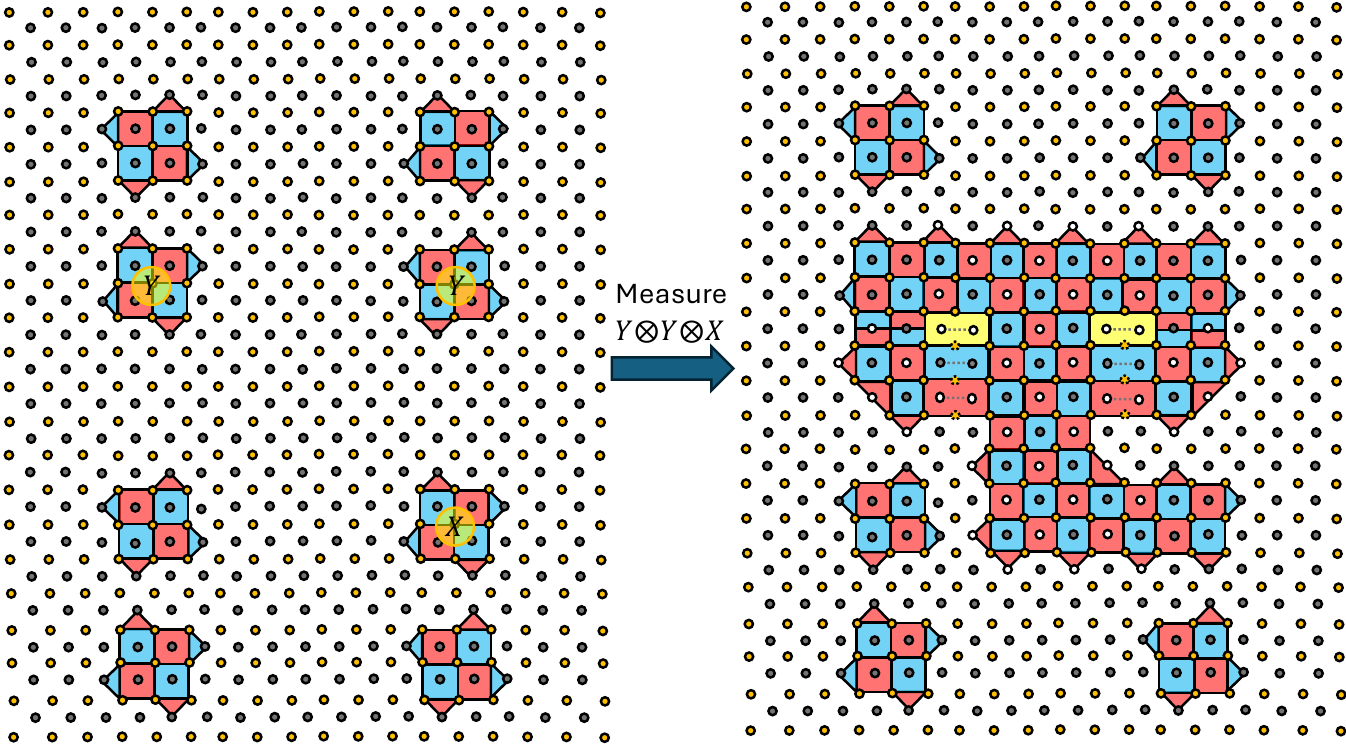}
        \caption{}
        \label{fig:LatticeSurgV1a}
    \end{subfigure}
    \hfill
    \begin{subfigure}[b]{0.7\linewidth}
        \centering
        \includegraphics[width=\linewidth]{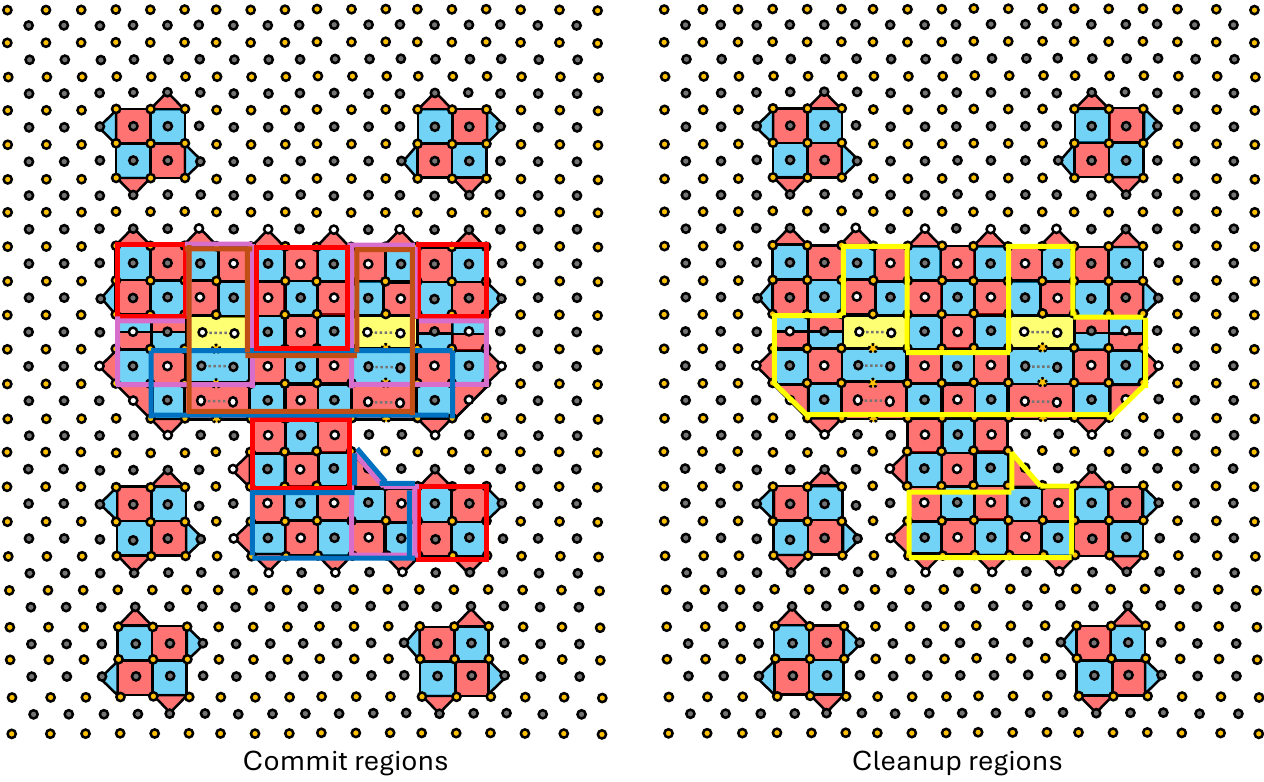}
        \caption{}
        \label{fig:LatticeSurgV1b}
    \end{subfigure}
    \caption{(a) Example of a logical $Y \otimes Y \otimes X$ measurement implemented via lattice surgery for surface code patches with $d=3$. The yellow vertices correspond to data qubits and grey vertices to ancillas. Red (blue) plaquettes correspond to $X$ ($Z$)-type stabilizers. Twist defects are represented by yellow plaquettes. White vertices represent random stabilizer measurements in the first syndrome measurement round when the patches are merged. However the product of all stabilizers with white vertices corresponds to the parity of the $Y \otimes Y \otimes X$ measurement. Stabilizer measurements of the merged surface code patch are measured over $d_m$ rounds, after which data qubits in the routing space region are measured in the $X$-basis which splits the surface code patches back to their original form. (b) Partitioning of a 2D slice of the merged lattice into commit (red boxes) regions and cleanup regions (yellow boxes). Errors in all commit regions are decoded in parallel. Likewise, errors in all cleanup regions are decoded in parallel. For commit regions, errors are corrected using stabilizer measurements from the red boxes, as well as surrounding buffer regions (shown by blue, purple and brown boxes).  }
    \label{fig:LatticeSurgFigs}
\end{figure}

Given access to magic states used as resource states, any universal quantum algorithm can be compiled in a sequence of multi-qubit Pauli measurements \cite{Litinski2019gameofsurfacecodes}. The fault-tolerant execution of multi-qubit Pauli measurements when using topological codes (such as the surface code) in hardware architectures constrained by nearest neighbor interactions requires a technique called lattice surgery \cite{2018arXiv180806709F,Litinski2019gameofsurfacecodes,Chamberland-2022-UniversalQuantumComputing}. In lattice surgery experiments, there can be logical spacelike failures (such as logical $X$, $Y$ or $Z$ errors occurring on logical qubits) or logical timelike failures, which occurs when the wrong parity of a multi-qubit Pauli measurement is obtained. Spacelike failures are exponentially suppressed with increasing code distance $d$, whereas timelike failures are exponentially suppressed with the number of syndrome measurement rounds $d_m$ performed during the measurement of a multi-qubit Pauli operator \cite{Chamberland-2022-UniversalQuantumComputing,PhysRevResearch.4.023090}. As such the runtime of an algorithm depends on both $d$ and $d_m$ (see for instance Eqs. (C7)-(C9) in Ref.~\cite{Chamberland-2022-UniversalQuantumComputing}). An example of a $Y\otimes Y \otimes X$ measurement is shown in \cref{fig:LatticeSurgV1a}.

In what follows, we assume that the depth of a quantum algorithm corresponds to the number of sequential non-Clifford gates (for instance $T$ gates or Toffoli gates). See Fig.6 in Ref.~\cite{Litinski2019gameofsurfacecodes} for an example (note that multiple non-Cliffords such as $T$ gates can be done in parallel, the depth of the circuit results from non-Clifford gates which cannot be parallelized). Before implementing the $j$'th sequential non-Clifford gate, the Pauli frame prior to the application of the gate must be known \cite{terhal2015,Chamberland2018faulttolerant}. As such, all syndrome measurement rounds up to the consumption of the magic state must be processed. As was shown in Ref.~\cite{Chamberland-2023-TechniquesCombiningFast-Preprint}, when decoding syndrome measurement rounds using a sliding window approach, the wait time needed to decode all syndrome measurement rounds for consumption of the magic state of the $j$-th sequential non-Clifford gate is
\begin{align}
    T^{b_j} = \frac{c^{j}r}{T_s^{j-1}} + T_l \Big[ \frac{T_s^{1-j}(c^j - T_s^j)}{c-T_s}  \Big],
    \label{eq:DecodeTimeBuffer}
\end{align}
where we assume a linear time decoder that takes time $T_{\text{DEC}}^{(r)} = cr$ to decode $r$ syndrome measurement rounds, and where $c$ is a constant. In \cref{eq:DecodeTimeBuffer}, $T_s$ is the time taken to perform one round of stabilizer measurements and $T_l$ is the round-trip communication latency of the Real-time Interconnect. For a PQPU with two-qubit gate times of $100 \text{ns}$ and a $1 \mu s$ measurement + reset time for ancillas, we can set $T_s = 1.4 \mu s$ for the surface code. If we assume that the number of syndrome measurement rounds prior to the consumption of the magic state used to implement the first non-Clifford gate is 33 (which applies to $d \approx 20$ surface codes when using lattice surgery to perform a $Z \otimes Z$ measurement) and take an inbound latency of $T_l = 20 \mu s$ (note that in this work, a $4 \mu s$ round trip latency was demonstrated), Ref.~\cite{Chamberland-2023-TechniquesCombiningFast-Preprint}, showed that $c \lesssim T_s$ results in wait times $T^{b_j}$ that grow sub-exponentially. The requirement that $c \lesssim T_s$ be challenging to achieve with realistic hardware and current state of the art QEC decoders. In Ref.~\cite{Skoric-2023-ParallelWindowDecoding,AlibabaParallelWindow}, an improvement was obtained by replacing a sliding window decoder with a parallel window decoder. Suppose that $L \ge r$ rounds need to be decoded for consuming a magic state and obtaining the current state of the Pauli frame. Instead of decoding the $L$ rounds in sequential batches of $r$ rounds, the $L$ rounds are partitioned into commit and buffer regions (where each contains $\mathcal{O}(d)$ rounds). Corrections in a given commit region using syndromes from the commit region as well as syndromes from buffer regions both immediately before and after the commit region. After performing such corrections, corrections in cleanup regions are performed to remove any remaining residual errors. See fig. 3 in Ref.~\cite{Skoric-2023-ParallelWindowDecoding}. Note that all commit and cleanup regions can be decoded in parallel which is a critical assumption for achieving real-time decoding in what follows. Let $N_{\text{par}}$ be the number of parallel resources, $n_{\text{com}}$ the number of rounds in a commit region and $n_W$ the number of rounds in a cleanup region. Let $2T_{\text{DEC}}$ be the time to decode the commit regions and cleanup regions. Ref.~\cite{Skoric-2023-ParallelWindowDecoding} showed that the exponential backlog problem can be avoided if $N_{\text{par}}$ satisfies
\begin{align}
    N_{\text{par}} \ge \frac{2 T_{\text{DEC}}}{(n_{\text{com}} + n_W)(T_l+T_s)}.
    \label{eq:Nparr}
\end{align}

The above discussion of parallel window decoding focused purely on the time domain. When performing lattice surgery, the effective distance $d_{\text{eff}}$ of merged surface code patches can be very large (see for instance \cref{fig:LatticeSurgV1a}). Decoding a surface code patch with $d_{\text{eff}} \gg d$ can result in $T_{\text{DEC}}$ times which are too large for practical considerations. However, commit/buffer and cleanup regions can also be used in the space domain in addition to the time domain. An example is provided in \cref{fig:LatticeSurgV1b}. In such a setting, commit regions and their corresponding buffer regions have both a spatial and temporal component and can all be decoded in parallel. Spatial cleanup regions correspond to residual portions of the lattice that need to be corrected after performing corrections in the commit regions. Similarly to commit regions, cleanup regions have both a spatial and temporal component and are also decoded in parallel. With enough parallel resources for spatial commit/cleanup regions, $T_{\text{DEC}}$ for lattice surgery can be made commensurate to $T_{\text{DEC}}$ times for pure memory settings with codes of distance $d$ even when $d_{\text{eff}} \gg d$.

We conclude this section by remarking that GPUs are particularly well suited for parallel window decoding protocols. Decoding can be viewed as an inference problem where each commit/buffer region corresponds to one element of the input batch size (and similarly for cleanup regions). In the context of AI-based decoders, if a GPU has enough memory to store the model, each batch element can (in principle) be processed in parallel. Although AI-based decoders are difficult to scale to large code distances \cite{AlphaQubitDecoder}, AI-based pre-decoders have been shown to scale to very large code distances \cite{Chamberland-2023-TechniquesCombiningFast-Preprint,Australia3DConv}. Pre-decoders correct most of the local error chains (which are the dominant sources of errors in topological codes) and thus can be used to significantly accelerate a global algorithmic decoder. Current work at NVIDIA is being done to build pre-decoders that can accelerate algorithmic decoders by more than an order of magnitude with the AI model having efficient implementations on a GPU \cite{AIpreDecNVIDIAV1}. Since modern GPUs can efficiently process large batch sizes, obtaining an $N_{\text{par}}$ that satisfies \cref{eq:Nparr} becomes more straightforward. 

\subsection{QPU Calibration}\label{scn:online-calibration}
Across the hardware stack of a future large scale quantum computer, there will be many compile-time and run-time decisions which must be made on-line without user intervention to decide things like the traversal direction of a calibration workflow given measured data, whether a particular characterization protocol should be mapped to a local control FPGA or out to a host CPU, and how to decode a set of error syndromes and with which computational resource. In particular, control systems with tightly coupled CPU and GPU compute will significantly improve the development and implementation of conditional calibration workflows which take advantage of in-depth characterization routines for the bring-up and maintenance of physical qubits in the presence of many sources of noise and time-dependent drift. 

Recent work details methods for bringing up a set of qubits from an uncalibrated state and then maintaining their performance using a conditional, directed, acyclic calibration graph~\cite{kelly2018}. In this premise, QPU developers design a calibration flow described by a graph structure that runs experiments, analyzes their measurement outputs to determine whether the parameters under test are within their desired specification, and then determines the next calibration step to run based on the current state of the system. The individual calibration steps can be simple, from sweeping a control frequency to determine a qubit’s resonance, to much more complicated routines which utilize characterization techniques like gate set tomography~\cite{nielsen2021gate} or randomized benchmarking~\cite{PhysRevA.77.012307, PhysRevLett.106.180504} to generate metrics which inform parameter updates. These characterization routines (discussed more thoroughly in Section \ref{scn:benchmarking}) can introduce much more computational complexity to design robust experiments, run complex circuits on the QPU quickly and efficiently, and fit their results. An architecture like the NVQLink which enables easier programming of these calibration graph structures will enable QPU development teams to bring up their devices faster and maintain high performance without having to invest as much overhead developing the firmware and software systems to enable this functionality. 

For simple calibration routines, the analysis required may involve simple peak detection or curve fitting, while more complex routines may require complex characterization. For example, running gate set tomography for even two qubits can require running thousands of quantum circuits along with computationally expensive optimization routines which take on the time scale of hours to solve on a desktop CPU. This generally disallows rapid feedback which would be useful for calibration. There have been demonstrations of hardware-accelerated implementations of smaller characterization routines like quantum state tomography in FPGA~\cite{miller2023}, but even this protocol runs into limits of a small FPGA’s LUT and BRAM resources for anything greater than six qubits. In practice, many research groups utilize faster but less informative characterization routines like randomized benchmarking (RB) to benchmark errors in their systems and determine if a calibration step improves or worsens the RB metric. For example, a laboratory experiment with CPU-hosted, AWG-based control for characterizing single qubit leakage rates using leakage RB~\cite{wallman2016} takes about ten minutes (dominated by circuit runtime), while running GST and extracting leakage rates for a single qubit takes on the order of an hour. Both circuit runtime and fitting the measurement results to useful metrics can be sped up by compute tightly coupled to the control system, enabling calibration routines which use more complex characterization for highly accurate parameter estimation. 

Additionally, stabilizing time-dependent drift is an important part of calibration workflows for quantum devices. Recent work has demonstrated single- and few-shot calibration protocols which operate very quickly to track and calibrate for drift in the parameter space of a quantum system~\cite{magannInPrep, miller2025}. These sorts of protocols can be integrated towards the end of calibration graph structures, where after a QPU has been brought to a stable operating point, low-latency calibration protocols take over to stabilize drift between running application circuits. However, even these intentionally lightweight protocols can require significant overhead depending on the control system implementing them and the number of qubits under calibration, and control systems must be able to implement corrections on a very fast time scale to track drift closely. For example, these protocols were demonstrated in an FPGA implementation and a cryogenic-CMOS ASIC simulation to be able to improve gate fidelities by orders of magnitude depending on the time between parameter updates~\cite{miller2025}. In this demonstration, the steady state infidelity of a simulated single-qubit gate with drifting optimal parameters reduced from roughly $10^{-4}$ to $10^{-5}$ when the calibration update time reduced from 10 microseconds to 1 microsecond, with further improvements achievable with faster updates. This would require fast pulse generation and classical compute which would be enabled by systems such as the NVQLink. The example in~\cite{miller2025} demonstrated the computational cost of tuning a single-qubit gate to be quite low, but scaling to the simultaneous drift control of hundreds to thousands of qubits and their gate operations will pose a challenge for the control system which would be aided by tightly coupled classical compute. Further, in the case of error-corrected logical qubits, the few-shot calibration protocols presented in~\cite{magannInPrep} can be extended to use syndrome data to inform calibration. This capability would be advanced by having the decoding systems and calibration systems present within the same tightly coupled control interface, such as in NVQLink. 

Overall, we see a wide variety of application spaces for low-latency calibration of quantum devices when tightly coupled with CPU and GPU compute. Scheduling calibration routines via an automated graph-like system, performing complex characterization routines in order to inform calibration steps, and maintaining low-latency communication for time-dependent drift control of a rapidly scaling parameter space all become significantly easier with tightly coupled compute which can quickly inform the control systems of the QPU. We see the NVQLink fitting well into this specific space to enable the creation of stable, scalable logical qubits.

\subsection{QCVV and benchmarking}\label{scn:benchmarking}
The computational power of quantum computers today is severely limited by noise and control errors. Noise characterization and calibration experiments are critically important to understanding this noise and maximizing these devices' capabilities. Quantum characterization, verification, and validation (QCVV) experiments, such as gate set tomography \cite{nielsen2021gate} or randomized benchmarking \cite{PhysRevA.77.012307, PhysRevLett.106.180504}, are valuable tools improving hardware performance, but are experimentally challenging and can easily outmatch the abilities of the classical control systems. These protocols can comprise hundreds or thousands of unique quantum circuits, can sometimes utilize random gates \cite{PhysRevLett.106.180504,wallman2016noise}, and can, independently, come with specifications to execute the circuits in strict order, returning time-stamped measurement results \cite{proctor2020detecting}. Circuits may need to be designed adaptively based on measurement results \cite{marceaux2023streaming} and dispatched to the QPU with low latency. Calibration experiments must further modify low-level pulse parameters \cite{rudinger2025heisenberg}, as frequently as after each circuit execution or even within individual circuits. Existing classical control infrastructure is often not up to the task. 

A major barrier to meeting these demanding requirements is the reliance on host-driven arbitrary waveform playback for pulse generation. In this conventional architecture, each control channel is driven by an arbitrary waveform generator (AWG) whose output is pre-computed on a host computer, uploaded to the instrument’s onboard memory, and played back verbatim during circuit execution. Because AWG memory is limited, the complete set of QCVV circuits often cannot be stored at once, requiring experiments to be divided into small batches. For each batch, all waveforms must be compiled, transferred to the instrument, and executed sequentially. This batching prohibits \emph{rastering} through the full circuit list, wherein one shot is taken from each circuit in turn, repeatedly, until the total number of desired shots is reached. Rastering of the data acquisition reveals time correlation and can reduce bias in error estimates due to drift, but batching makes it impossible. These disadvantages are compounded by the lengthy upload times involved. Transferring waveforms for batches of circuits, which can number in the hundreds or thousands, takes anywhere from minutes to hours. Such delays not only incur significant quantum downtime but also fragment the experimental process, making it difficult to capture precise snapshots of device performance at a particular point in time.

Table \ref{tab:QCVV_resource_requirements} summarizes the controller resource requirements for four classes of QCVV experiment: traditional ``static" QCVV, adaptive QCVV, single-shot calibration, and calibration using deep reinforcement learning. Each stresses different combinations of the controller subsystems. 

\begin{table}[htbp]
	\centering  
	\scalebox{.8}{
		\begin{tabular}{c|cccccc}
			\hline
			\multirow{2}[2]{*}{Application} & \multicolumn{6}{c}{Resource Requirements} \bigstrut[t]\\
			& \makecell{Controller\\Memory} & \makecell{Computational\\Throughput} & \makecell{Latency\\(Real-time\\ Interconnect)} & \makecell{Latency\\(Host\\Connection)} & \makecell{Real-time\\Pulse Updates} & \makecell{Control Flow\\Flexibility} \bigstrut[b]\\
			\hline
			%    Traditional ``Static" QCVV & 3     & 1     & 1     & 2     & 0     & 2 \bigstrut[t]\\
			%    Adaptive/Online QCVV & 3     & 3     & 3     & 2     & 1     & 3 \\
			%    Calibration (Single-Shot) & 1     & 2     & 3     & 1     & 3     & 2 \\
			%    Calibration (RL) & 3     & 3     & 3     & 1     & 3     & 3 \bigstrut[b]\\
			Traditional ``Static" QCVV & High     & Low     & Low     & Medium     & None     & Medium \bigstrut[t]\\
			Adaptive/Online QCVV & High     & High     & High     & Medium     & Low     & High \\
			Calibration (Single-Shot) & Low     & Medium     & High     & Low     & High     & Medium \\
			Calibration (RL) & High     & High     & High     & Low     & High     & High \bigstrut[b]\\
			\hline
	\end{tabular}}%
	\caption{Classical controller resource requirements for four representative classes of QCVV experiments/protocols. Different classes of QCVV protocol can vary widely in their overall resource requirements and the specific subsystems they stress. %(0:None, 1:Low, 2:Medium, 3:High)
	}
	\label{tab:QCVV_resource_requirements}%
\end{table}%

Traditional ``static" QCVV protocols include common experiments such as randomized benchmarking and gate set tomography (GST) and involve the construction of experiments consisting of possibly hundreds or even thousands of circuits specially structured to learn one of more properties of a quantum processor, \textit{e.g.}, gate or SPAM infidelities, with most computational analysis performed on a host system in post-processing. While these experiments don't rely heavily on access to realtime compute resources, they very frequently stress available memory resources, particularly in older AWG-based controllers. Real-time controllers based on FPGAs or RFSoCs are able to generate waveforms on the fly based on few-bit keywords. In these cases, increased flexibility in control workflows significantly reduces both controller memory requirements and the latency requirements with a host system \cite{xu2023qubic}.

Adaptive QCVV experiments---ones which aim to learn the characteristics of a quantum processor in real-time with dynamically constructed experiment designs---and real-time calibration protocols demand much more from the control hardware. They require nearly real-time interconnects between the controller and host, to enable shot-by-shot updates to control parameters, or even within-shot updates. Even slight delays can hinder the performance of feedback loops and complicate the execution of branching control logic. Meanwhile, the computational throughput needed for on-the-fly decision making can vary widely among these tasks. For instance, simple fast-feedback routines may require only basic arithmetic with minimal memory, but when controllers integrate complex algorithms, such as those based on deep reinforcement learning or requiring decoding of QEC syndrome data, they increasingly stress every aspect of the control system.

Modern real-time controllers based on FPGAs or RFSoCs integrate digital waveform synthesis and sequencing directly in hardware. Rather than uploading large waveform buffers, they generate parameterized pulses in real time, enabling adaptive control, feedback, and fine-grained synchronization without the bottlenecks of host-side compilation and transfer, or the memory requirements of storing precompiled waveforms. This increased flexibility in control workflows significantly reduces both controller memory requirements and the latency requirements with a host system \cite{xu2023qubic}. Further incorporating low-latency interconnects to fast classical co-processing can enable even the most demanding adaptive calibration and characterization routines. Tiered heterogeneous control architectures, like NVQLink, are wellsuited to meet these needs by offloading light computational tasks, such as basic arithmetic or inference on small neural networks, to FPGA hardware for rapid processing. More complex tasks, including training and inference of statistical models or larger neural networks, are handled on CPU or GPU nodes via a slightly higher-latency connection.

Transitioning away from host-controlled AWG controllers in favor of these lower-latency architectures will not just allow existing QCVV and calibration routines to run as intended, but will also encourage development and deployment of new classes protocols that will enable hardware developers and users to probe and mitigate errors with ever finer resolution and speed.

\section{Development and Simulation}\label{scn:development}

Although the primary motivation for the \nvqlink architecture is to support quantum computing at scale, we recognize that the utility of any platform depends strongly on the ability to develop and maintain software on that platform.
This section describes offline tools that support quantum program development and validation
outside the real-time execution path, enabling developers to test and refine quantum programs before
deployment on physical hardware.

CUDA-Q~\cite{cudaq} provides logical-level circuit simulators (state vector, density matrix, tensor network)~\cite{cuquantum} to support research in quantum computing and application development~\cite{cudaqx}.
Beyond this logical layer, we distinguish two complementary capabilities
for physical hardware emulation: the VPPU, which emulates PPU
instruction-level behavior to enable offline testing of compiled quantum programs, and PQPU simulators, which model quantum dynamics at the Hamiltonian level to provide insights
into physical implementation fidelity. These tools operate at different architectural layers---VPPU
at the PPU compiler and instruction set architecture (ISA) level, and PQPU simulators at the quantum state
evolution level.
The VPPU abstraction is introduced in~\autoref{scn:vppu}, and we include a few notes on the role of PQPU simulation in~\autoref{scn:physical-qpu-sim}.
A particularly important use case in focus here is the simulation of QEC encoded programs, which we discuss in~\autoref{scn:qec-sim}.

\subsection{Virtual Pulse Processing Unit}\label{scn:vppu}
Developing and validating quantum programs for tightly coupled systems presents a fundamental challenge: compiled programs must target PPU-specific instruction sets with precise timing constraints, yet validating these programs traditionally requires deploying to physical hardware.
This creates development bottlenecks, as iterations on quantum control logic, real-time protocols (QEC decoding, adaptive calibration), and pulse sequences require continuous hardware access.
The VPPU addresses this challenge by providing a substitutable PPU emulator that enables offline program validation in the PTD. A recent prototyping attempt \cite{ye2025emuplat} has shown promise in achieving full pipeline validation. 
By transforming ISA instructions into an appropriate signal representation for physical QPU simulation (eg. $V(t)$ in the control Hamiltonian of the PQPU, where $t$ evolves in PTD), the VPPU allows developers to test instruction sequences, validate timing constraints, and debug pulse schedules before committing to physical execution, thereby accelerating development cycles for both quantum programs and real-time classical protocols.

As defined in ~\autoref{scn:system-components}, the VPPU must be substitutable for the
physical PPU. This substitutability ensures that quantum programs compiled for physical hardware can be tested and validated offline
without modification. VPPU implementations achieve this by implementing the \literal{quantum_control_trait}
interface defined in the runtime architecture (~\autoref{scn:devices}), presenting the
same programmatic interface as their corresponding physical PPU devices.

If offered, VPPU implementations must:
\begin{itemize}
    \item Fully implement target PPU's instruction set (ISA)
    \item Transform ISA control instructions to signal representations usable in PQPU dynamical simulation
    \item Emulatie PQPU readout: convert signal representations in a PQPU simulation into correct PPU data formats
    \item Maintain compatibility with the PPU compiler
    \item Support the same trait interface as its physical PPU or QSC
\end{itemize}

The signal representations produced by VPPU enable offline inspection and visualization of compiled
pulse sequences. Developers can analyze timing relationships, identify potential resource
conflicts, and validate pulse scheduling logic before deploying to physical hardware.
This capability is particularly valuable for debugging complex multi-qubit operations
and verifying that compiled programs meet timing constraints of the target QPU modality.

VPPU implementations maintain an interface that transforms ISA instructions
into signal representations in the PTD. To achieve this output, implementers may
choose various approaches---from lookup tables for simple waveforms to full quantum
dynamics simulation. Some VPPU implementations may use Physical QPU simulators
(\autoref{scn:physical-qpu-sim}) as computational backends for computing eg. $V(t)$ from
ISA instructions, but this is entirely an implementation choice invisible to calling code.
The VPPU interface remains strictly at the PPU instruction level (ISA $\rightarrow V(t)$),
maintaining clear architectural separation from Hamiltonian-level simulation.
Higher-level optimization tasks such as reinforcement
learning-based calibration (\autoref{scn:online-calibration}), or QEC decoder training
(~\autoref{scn:qec-sim}) are independent tools that may use VPPU as a library component.

\subsection{Physical QPU Simulation}\label{scn:physical-qpu-sim}

Physical QPU simulators are computational tools that model quantum
dynamics at the Hamiltonian level, often operating on quantum state vectors or density matrices.
While a VPPU emulates
PPU instruction execution (ISA $\rightarrow V(t)$), a PQPU simulator models the fundamental
quantum state evolution (Hamiltonian $\rightarrow |\psi\rangle$), where the time variable
$t$ in the control Hamiltonian evolves in the PTD.

PQPU simulators can function
as backends for VPPU implementations to provide quantum state fidelity modeling under
physical noise, as tools for physical parameter characterization and optimization, or as
While they need not run in the RTD as realtime applications, we expect that in practice they will usefully be supported on the Real-time Host in a timesharing mode.

\subsection{QEC simulation}\label{scn:qec-sim}
Simulation of quantum programs is fundamental to quantum algorithm research and development, and this remains true in the study of quantum error correction (QEC).
However, QEC protocols impose unique demands on simulation, emphasizing different dimensions compared to traditional algorithmic workloads.
Like quantum algorithms, simulating QEC routines allows researchers to validate and optimize new proposals.

A key focus in these studies is the logical error rate of a specific protocol.
For instance, a quantum memory experiment may involve preparing a logical quantum state and performing repeated rounds of error correction to extend the logical qubit’s lifetime beyond that of its constituent physical qubits.
Such processes can be simulated, where the resulting logical error rate depends on three main factors: the quantum circuit, the noise model, and the decoder.
Understanding how each of these influences logical qubit fidelity is a central task for researchers seeking to design the next generation of fault-tolerant quantum architectures.

Although the utility of QEC simulation is considerable, the structure of these protocols poses challenges for conventional approaches.
Because a logical qubit is encoded into many physical qubits, QEC simulation scales poorly with techniques such as density matrix (DM) or statevector (SV) simulation.
Tensor network (TN) simulators can handle larger qubit counts when entanglement remains sparse, but QEC circuits typically employ multiple layers of two-qubit gates, quickly reaching the limits of TN efficiency.
Moreover, frequent mid-circuit measurements common in QEC workflows further complicate these simulation methods.

One major opportunity lies in the fact that much of QEC involves Clifford operations, which can be simulated efficiently using stabilizer simulators such as Stim~\cite{Gidney2021stimfaststabilizer}.
Stim enables researchers to explore the interplay of circuits, noise models, and decoders: three essential components of QEC studies.
Its design excels for offline decoding workloads, where large batches of simulated shots are generated and later decoded to evaluate protocol performance.
Stim’s circuit language is optimized for this case and prioritizes execution speed, though it does not support conditional gate application.

The gap across these simulation approaches is the inability to model full QEC workflows, including scenarios such as magic state distillation and conditional gate execution based on mid-circuit decoding results.
Addressing this gap is a key motivation behind the NVQLink architecture, which supports such workflows through the \literal{cudaq::device_call} interface.
Library code can be written to switch seamlessly between real QPU execution and an emulated mode for simulation.
\begin{lstlisting}[
    language=C++,
    caption={Enqueue syndromes wrapper function.},
    label={lst:qec-enqueue},
    float=htbp
]
__qpu__ void
enqueue_syndromes(std::uint64_t decoder_id,
                  const std::vector<cudaq::measure_result> &syndromes,
                  std::uint64_t tag) {
#ifdef LIBSIM
  cudaq::device_call(enqueue_syndromes_simulation, decoder_id, syndromes, tag);
#else
  uint64_t syndrome_size = syndromes.size();
  uint64_t syndrome = cudaq::to_integer(syndromes);
  cudaq::device_call(enqueue_syndromes_ui64, decoder_id, syndrome_size,
                     syndrome, tag);
#endif
}
\end{lstlisting}

In~\autoref{lst:qec-enqueue}, we show an example of sending measurement data to a decoder.
When the library is compiled with \literal{-DLIBSIM}, the \literal{enqueue_syndromes} function produces a binary optimized for simulation, eliminating the need to pack measurement data into integers by directly using the size field of the \literal{std::vector}.
When targeting real hardware, the developer instead packs the measurement results explicitly and specifies the number of syndrome bits.
Both cases are represented in the example, enabling the same application code to be validated under simulation before being re-targeted to an experimental control system.

By incorporating decoder calls and conditional gate logic directly in application code, this design also generalizes across simulation strategies.
For example, extended stabilizer simulation for scenarios which are very nearly Clifford, or back to state vector simulation when qubit counts are low but advanced noise modeling is needed.
Stabilizer-based simulation will serve as a natural starting point for many QEC workloads, yet CUDA-Q kernels written in this style can retarget to any simulation strategy, or hardware backend, so long as the necessary data transfer semantics are defined via the appropriate \literal{cudaq::device_call} bindings.

\section{Conclusion}\label{scn:conclusion}

We have presented \nvqlink, a platform architecture that tightly couples high-performance classical compute to quantum-processor control systems.
The architecture defines a model of a Logical QPU comprising CPUs, GPUs, and PPUs connected by a low-latency Real-time Interconnect, and it distinguishes four time domains —- Physical (PTD), Deterministic (DTD), Real-time (RTD), and Application (ATD) -— to reason about correctness, latency budgets, and other performance requirements.
A RoCE-based proof of concept demonstrates sub-\qty{4}{\,\micro\second} steady-state round-trip latency with commodity networking equipment and a GPU-resident loopback path, indicating a practical route to scaling radix and bandwidth while keeping jitter low.

We presented proposed extensions to CUDA-Q with device-addressable real-time callbacks (\texttt{cudaq::device\_call}) and a heterogeneous memory abstraction (\texttt{cudaq::device\_ptr<T>}), enabling quantum kernels to invoke real-time accelerated computation with compiler-managed marshaling.
We outlined compilation strategies across latency regimes—AOT lowering to pulse/ISA with RDMA and persistent kernels for short-timescale modalities, and host-mediated streaming with JIT for longer-timescale modalities—expressed via Quake/CC MLIR and unified lowering.

Our proposed runtime architecture offers a zero-overhead, trait-based model to compose device capabilities—explicit data marshaling, device callbacks, quantum control (\literal{quantum_control_trait}), and RDMA—plus representations for compiled quantum kernels, pluggable compilers/executors, and a Logical QPU Driver API for allocation, transfer, and launching on both physical PPUs and substitutable VPPUs.

We surveyed QPU-level workloads, including an example of $T$ state distillation, showing how GPU batch inference and pre-decoders reduce backlog and align with RTD constraints. For calibration and QCVV, we showed how moving beyond host-driven AWG playback to parameterized, real-time control with GPU/CPU co-processing enables adaptive protocols, short-cycle drift mitigation, and within-shot feed-forward.

Finally, we described development tools: VPPU as a drop-in emulator at the PPU ISA boundary, PQPU simulators for Hamiltonian-level studies, and CUDA-Q simulators for logical-level testing—allowing the same source to retarget between emulation and hardware via \texttt{device\_call}.

We invite QPU and QSC builders, HPC centers, and researchers to evaluate and engage with our proposal and communicate with us about requirements.
Our intent is a pragmatic, open path to real‑time accelerated computing in the QPU domain that scales from today's devices to fault‑tolerant systems.

\clearpage

\begin{acks}
The authors would like thank the following individuals for helpful conversations and feedback in preparing the material for this manuscript:
Owen Arnold,
Simeon Baker-Finch,
Francesco Battistel,
Gilad Ben-Shach,
Matthew Bradley,
Gustavo Cancelo,
Sergio Cantu,
Arnaud Carignan-Dugas,
Yonatan Cohen,
Coleman Collins,
Niccolo Coppola,
Erik Davis,
Patrick Deuley,
Nicolas Didier,
Mengke Feng,
Brett Freeman,
Louis Fry-Bouriaux,
Joanna Fulton,
Pranav Gokhale,
Eric Holland,
Andrew Houck,
Matthew Hutchings,
Jerome Javelle,
Ronan Jezequel,
Gwen Johnson,
Glenn Jones,
Ryan Jones,
William Kindel,
Shuyi Liu,
Jie Luo,
Jeffrey Marshall,
Josh Moles,
Yasunobu Nakamura,
Tom Noel,
Thomas Ohki,
Simon Philips,
Laurent Prost,
Matthew Reagor,
David Reilly,
Hengjiang Ren,
David Rivas,
Yoav Romach,
Christoph Rühle,
Colm Ryan,
Kentaro Sano,
Andre Saraiva,
Mitsuhisa Sato,
Laura Schulz,
Michael Sorenson,
Ramon Szmuk,
Ryousei Takano,
Brian Tarasinski,
Jeff Thompson,
Sho Uemura,
Oded Wertheim,
Evan Zalys-Geller,
David van Zanten,
Avishai Ziv.

Support is also acknowledged from the U.S. Department of Energy, Office of Science, National Quantum Information Science Research Centers, Quantum Systems Accelerator under Air Force Contract No. FA8702-15-D-0001. These results are based in part upon work supported by the U.S. Department of Energy, Office of Science, National Quantum Information Science Research Centers, Quantum Science Center. Any opinions, findings, conclusions, or recommendations expressed in this material are those of the author(s) and do not necessarily reflect the views of the Dept. of Energy. Y.J. would like to thank the National Research Foundation, Singapore, National Quantum Office under
its National Quantum Computing Hub and Hybrid Quantum-Classical Computing 1.0 programmes. A*STAR under C23091703 and Q.InC Strategic Research and Translational Thrust for the funding support. Y.X., G.H., I.S. would like to thank the support from the U.S. Department of Energy, Office of Science, Advanced Scientific Computing Research (ASCR) Quantum Testbed Program under Contract No. DE-AC02-05CH11231.

Sandia National Laboratories is a multimission laboratory managed and operated by National Technology \& Engineering Solutions of Sandia, LLC, a wholly owned subsidiary of Honeywell International Inc., for the U.S. Department of Energy’s National Nuclear Security Administration under contract DE-NA0003525. This paper describes objective technical results and analysis. Any subjective views or opinions that might be expressed in the paper do not necessarily represent the views of the U.S. Department of Energy or the United States Government.
\end{acks}

%% Bibliography
% \bibliographystyle{ACM-Reference-Format}
\bibliographystyle{unsrtnat}
\bibliography{main}

\clearpage
\appendix
% \printglossary[title=Definitions, toctitle=Defintions]

\end{document}